\documentclass[a4paper,onecolumn,11pt,unpublished]{quantumarticle}
\pdfoutput=1
\usepackage[utf8]{inputenc}
\usepackage[english]{babel}
\usepackage[T1]{fontenc}
\usepackage{amssymb,amsthm,amsmath}
\usepackage{nccmath}
\usepackage{hyperref}
\usepackage[numbers,sort&compress]{natbib}

\usepackage[table,svgnames]{xcolor}

\usepackage{tikz}
\usepackage{lipsum}

\usepackage{mathtools}
\DeclarePairedDelimiter\bra{\langle}{\rvert}
\DeclarePairedDelimiter\ket{\lvert}{\rangle}
\DeclarePairedDelimiterX\braket[2]{\langle}{\rangle}{#1\,\delimsize\vert\,\mathopen{}#2}

\newcommand{\y}[1]{\,\mathrm{#1}} 
\newcommand\identity{1\kern-0.25em\text{l}} 
\usepackage{ytableau} 

\begin{document}

\title{Group-theoretic treatment of strong light--matter coupling with an arbitrary number of excitations}

\author{Antti Peltola}
\affiliation{Department of Mechanical and Materials Engineering, University of Turku, FI-20014 Turku, Finland}
\orcid{0009-0007-9667-3881}

\author{Olli Siltanen}
\email{olmisi@utu.fi}
\affiliation{Department of Mechanical and Materials Engineering, University of Turku, FI-20014 Turku, Finland}
\orcid{0000-0002-7295-2065}

\author{Kimmo Luoma}
\email{ktluom@utu.fi}
\affiliation{Department of Physics and Astronomy, University of Turku, FI-20014 Turku, Finland}

\author{Konstantinos S. Daskalakis}
\affiliation{Department of Mechanical and Materials Engineering, University of Turku, FI-20014 Turku, Finland}
\orcid{0000-0002-3996-5219}

\maketitle

\begin{abstract}
Strong light--matter interactions in optical microcavities give rise to hybrid light--matter states known as polaritons. While actively used in modern technologies, theoretical descriptions of such systems are often restricted to the single-excitation case, limiting their ability to capture many-excitation physics and hindering further technological advancements. Here, by exploiting the combinatorial structure of quantum emitters, we investigate the Tavis-Cummings model with arbitrary number of excitations. We derive the structure and properties of its eigensystem and identify allowed radiative transitions in systems of realistic size scales. Our work reveals new behavior inaccessible to the few-excitation regime, while also providing a framework to reduce the computational complexity of similar systems with exponentially growing Hilbert spaces.
\end{abstract}

\section{Introduction}

\begin{figure}[t]
  \centering
  \includegraphics[width=.95\linewidth]{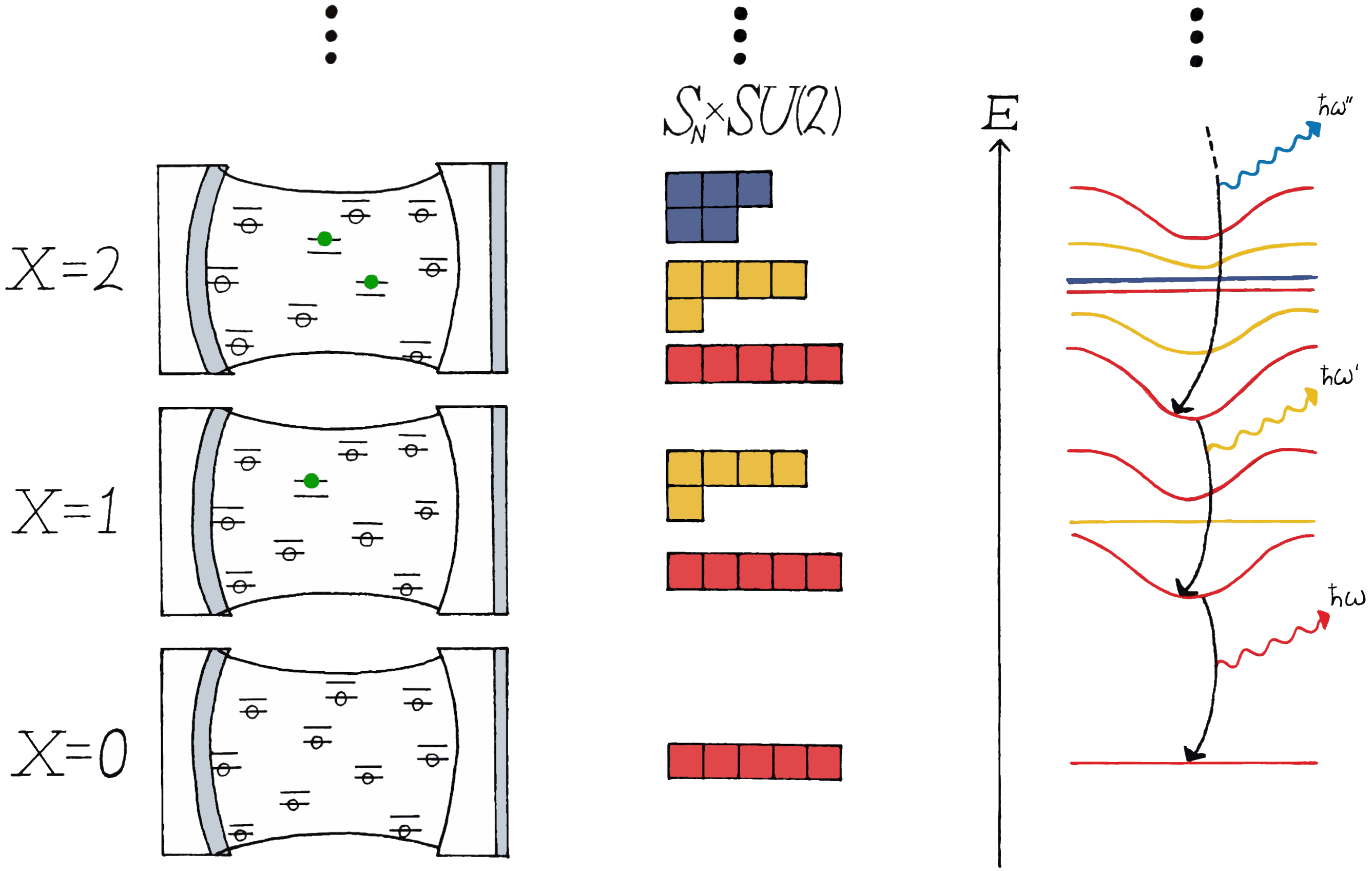}
  \caption{Schematic illustration of the system, model, and predictions. The Hilbert space corresponding to the physical system (two-level systems and cavity mode) grows exponentially with the excitation number $X$.
  This growth, however, remains computationally manageable when exploiting the algebraic structure of the Hilbert space. The structure, visualized here by the Young tableaux, manifests itself as symmetries in the eigenenergies (dispersion curves) and observable transition energies (wavy arrows) of the system.}
  \label{fig:figure1}
\end{figure}

Strongly interacting dipolar quantum emitters and confined electromagnetic field modes form hybridized energy states known as polaritons, which have long stood as the center of attention of modern fundamental and material research~\cite{Qureshi2026, Polaritonic_quantum_matter, Siltanen_MH, Molecular_polaritons_review, Byrnes2014, Bhuyan2023, Lednev2024, Sandik2024, Dutta2024}. Polariton studies shed light between the quantum and classical understanding of physics while simultaneously paving the way for novel applications, e.g., in organic optoelectronics and energy storage~\cite{Sanvitto2016,Hymas2026,Mischok2024,Abdelmagid2025}. Systems of $N$ emitters are often modeled by the Dicke model, Tavis-Cummings (TC) model, and their extensions~\cite{Dicke,Tavis-Cummings}. These have been experimentally verified at low excitation numbers~\cite{Haroche_experiment}, in which case they are also exactly solvable~\cite{Skrypnyk_2018,PAN200594}. However, as the number of operations grows exponentially with realistic numbers of emitters ($N$) and excitations ($X$), the system becomes practically unsolvable~\cite{Campos-Gonzalez-Angulo_1}.

Symmetries have provided a strong tool to characterize light--matter interactions~\cite{Braak,Vandaele_2017,Skrypnyk_2018}. Some modern attempts to solve the many-excitation problem have used permutational symmetries of the system, removing the $N$-dependence from the computational task~\cite{Campos-Gonzalez-Angulo_1,Campos-Gonzalez-Angulo_2,Sanchez_CUT-E}. Along with the well-known decomposition of the Hilbert space into excitation manifolds defined by fixed $X$~\cite{Campos-Gonzalez-Angulo_1}, diagonalizing the system is reduced to solving a set of matrices with dimensions $\mathcal{O}(X)$, in cardinality of the same order. This approach has been used to probe the structure and dynamics of manifolds beyond the ubiquitous $X=1$ restriction~\cite{Siltanen_AOM,Borges_Dark_Polariton,Campos-Gonzalez-Angulo_2}.

Recent applications of light--matter interactions have considered consumer applications and systems closer to macroscopic scale, motivating the study of these more realistic regimes~\cite{Pandya2021,Tibben_2025}. It is in this regime where theoretical understanding of model systems has not yet caught up with experimental studies~\cite{Mandal,Fregoni2022-qg,Weight_ab_initio_methods}. Interestingly though, models restricting to $X=1$ seem to work well for predicting energy transitions in polaritonic systems. Here we provide an interpretation of why this is by modeling the nonlinear regime of more excitations~\cite{Ribeiro_nonlinear}. Energy spectra and optical dynamics gain anharmonic effects at these regimes.

In this paper, we extend the polariton theory to an arbitrary number of emitters and excitations by taking advantage of symmetry found in the system (see Fig.~\ref{fig:figure1}). The rotating-wave approximation (RWA) makes the total excitation number a conserved quantity, splitting the Hilbert space into finite-dimensional excitation manifolds~\cite{Tavis-Cummings}. Simultaneously, permutational action of the symmetric group divides the space into irreducible representations (irreps), removing direct dependence on the number of emitters from the computational load of solving the system~\cite{Campos-Gonzalez-Angulo_1, Fulton1991RepresentationTA}. Under uniform coupling, this allows us to algorithmically solve the matrix spectrum and eigenstates of the system for arbitrary system parameters $N$ and $X$.

We first establish the fundamental nature of this extension of the theory, setting the stage for further studies in directions of case-by-case motivations. We then apply the theory to find selection rules for the dynamical generators of a lossy cavity. These rules allow us to predict emissive behavior of the system, such as emission from the lowest polariton blue-shifting with increasing $X$. We are also able to discuss why the well-established $X=1$ theory works so well with experimental results.

\section{The model}\label{sect:Structure_theory}

\subsection{Structure theory}

The TC Hamiltonian describing a system of $N$ two-level systems (TLSs) coupled uniformly to a single electromagnetic mode  can be written under the RWA as $H_{\mathrm{TC}} =H_0 + H_I$, where
\begin{align} 
     H_0 &=
     \sum_{n=1}^N E_{TLS} \ket{e_n}\bra{e_n}
    + E_c\hat{a}^{\dagger}\hat{a}, \label{eqn:free_Hamiltonian}  \\
    H_I&=g \left( \hat{S}^{\dagger}\hat{a} + \hat{S}\hat{a}^{\dagger} \right), \label{eqn:interaction_Hamiltonian}
\end{align}
with $\hat{S} := \sum_{n=1}^N \hat{\sigma}_-^{(n)}$ being the collective spin lowering operator for $N$ TLSs. Under the RWA, the Hamiltonian commutes with the total excitation number operator $\hat{X} := \hat{S}^{\dagger}\hat{S} + \hat{a}^{\dagger}\hat{a}$, meaning that the total excitation number $X$ is conserved. This allows one to separate the Hilbert space into an infinite sum of finite-dimensional excitation manifolds according to the eigenvalue $X$, as
\begin{align} 
    \mathcal{H} &=\mathcal{H}^{(0)} \oplus \mathcal{H}^{(1)} \oplus \mathcal{H}^{(2)} 
    \oplus \cdots \oplus \mathcal{H}^{({X_{\y{max}}-1})} \oplus \mathcal{H}^{({X_{\y{max}}})} \oplus \mathcal{H}^{({X_{\y{max}}+1})} \oplus \cdots,
\end{align}
where in reality the occupied Hilbert space is limited by some $X_{\y{max}}$ as the energy of a physical system must be finite. This corresponds to a block-diagonal form of the Hamiltonian, known to have corresponding Hilbert space dimensions

\begin{align}
    H_{\mathrm{TC}} &= \bigoplus_X H ^{(X)}, \\
    \mathrm{dim}~\mathcal{H} &= \sum_{l=0}^{X_{\text{max}}} \mathrm{dim}~\mathcal{H}^{(l)}
    = \sum_{j=0}^{X_{\text{max}}} \sum_{j=0}^i \binom{N}{j},\label{eqn:manifold_dimension}
\end{align}
up to the maximum excitation number considered~\cite{Borges_Dark_Polariton}. 

The system can be given more structure by taking note of the symmetries of the Hamiltonian. The Hilbert space can be decomposed into photonic (Fock space) and material (TLS) parts as $\mathcal{H} = F \otimes \left( \mathbb{C}^2\right)^{\otimes N}$. This allows for a natural action of the Lie algebra $\mathfrak{su}(2)$ on the individual TLS spaces, and of the symmetric group $S_N$ on the tensor power of TLSs. One can then decompose the system into irreps of the symmetric group. With uniform coupling constant $g$, the Hamiltonian is invariant under the action of $S_N$ permuting the TLSs, making the closed dynamics respect this decomposition. Since the actions of the two groups commute, one can apply the well-known \textit{Schur-Weyl duality} from representation theory, resulting in these irreps combining with the ones of $SU(2)$ under joint labels $\lambda$, which correspond to Young diagrams of $S_N$ or equivalently weights (spins) $S$ of the Lie group $SU(2)$. The \textit{cooperation number} or \textit{total spin} $S$ is the usual eigenvalue of the Casimir operator $\hat{S}^2$ of $\mathfrak{su}(2)$ \cite{Klimov_Chumakov}. The ``2'' of $SU(2)$ limits the number of rows of the considered Young diagrams (Appendix~\ref{app:Young_tableaux}) so that we can write $\lambda = (N-m,m)$, where $m$ is the number of cells on the second row in English notation~\cite{Sagan}. Because a Young diagram cannot have more cells on the second row than the first row, we find that $0\leq m\leq N/2$ (for odd $N$, $0 \leq m \leq \left \lfloor N/2 \right\rfloor$). Each excitation manifold then further decomposes into irreps as

\begin{equation} 
    \mathcal{H}^{(X)}
    \stackrel{\mathclap{\tiny\mbox{S-W}}}{\cong}
    \bigoplus_m \left( \pi_m \otimes W_m \right)
     =
    \bigoplus_m \left(\mathcal{H}_m^{(X)} \right)^{\bigoplus d(N,m)} \equiv \bigoplus_m  V_m^{(X)},
\end{equation}
where $\pi_m$ is an irrep of $S_N$, $W_m$ is an irrep of $SU(2)$, $m = \frac{N}{2}-S$ connects the Young diagrams and weights of $SU(2)$, and $d(N,m)$ is the multiplicity of the irrep $\mathcal{H}_m^{(X)}$, given by the dimension of the corresponding irrep of the symmetric group $\pi_m$~\cite{Sagan}. The block-diagonal TC Hamiltonian of each manifold can then be written as

\begin{align} 
    H^{(X)} &= \bigoplus_m H_{V_m}^{(X)} 
    = \bigoplus_m \left( \identity_{d(N,m)} \otimes H_m^{(X)} \right) \notag \\
    &=
    \begin{psmallmatrix}
        \text{\normalsize$H_0^{(X)}$}&&&&&&&&&&&\\
        &\text{\normalsize$\ddots$}&&&&&&&&&&\\
        &&\text{\normalsize$H_0^{(X)}$}&&&&&&&&&\\
        &&&\text{\normalsize$H_1^{(X)}$}&&&&&&&&\\
        &&&&\text{\normalsize$\ddots$}&&&&&&&\\
        &&&&&\text{\normalsize$H_1^{(X)}$}&&&&&&\\
        &&&&&&~~~&&&&&\\
        &&&&&&&\text{\normalsize$\ddots$}&&&&\\
        &&&&&&&&~~~&&&\\
        &&&&&&&&&\text{\normalsize$H_X^{(X)}$}&&\\
        &&&&&&&&&&\text{\normalsize$\ddots$}&\\
        &&&&&&&&&&&\text{\normalsize$H_X^{(X)}$}
    \end{psmallmatrix}, \label{eqn:manifold_block_decomposition}
\end{align}
where each submatrix $H_m^{(X)}$ is repeated $d(N,m)$ times. Due to the combinatorial nature of the symmetric group, the integer $m$ is further limited by the number of excited TLSs being permuted among ground-state TLSs. The maximum value of $m$ within an excitation manifold is therefore

\begin{align}   
    m_{\y{max}} &= \min\{N/2, X\}.
\end{align}

\subsection{Dimension formulas}

The dimensions of the irreps $\mathcal{H}_m^{(X)}$ follow the integer dimension of the weight spaces of $SU(2)$, given by $X$ and $m$ as

\begin{align}\label{eqn:irrep_dimension}   
    \dim{\mathcal{H}_m^{(X)}} &= \min\{ X-m+1, N-2m+1\},
\end{align}
where $N-2m+1 = 2S+1$, using the composite spin quantum number $S$. If $m>X$, $\dim{\mathcal{H}_m^{(X)}} := 0$. Since $N-2m+1 = 2S + 1$ is the dimension of the $SU(2)$ irrep of weight $S$, it must be the maximal dimension of each $\mathcal{H}_m^{(X)}$. The irreps are then repeated $d(N,m) = \dim{\pi_m}$ times. The dimension of the $S_N$ irrep $\pi_m$ is famously given by the Hook length formula, a combinatorial formula calculated using ``hook lengths'' of a Young diagram of shape $\lambda$ (see Appendix \ref{app:Young_tableaux}). For two rows, this formula can be written analytically as

\begin{align}\label{eqn:multiplicity_formula}   
    d(N,m) &= 
    \binom{N}{m-1}\frac{N-2m+1}{m}.
\end{align}
Combining the last two equations and writing out the binomial coefficient, one gets the dimension formula
\begin{equation} \label{eqn:dimension_formula}
    \mathrm{dim}~V_m^{(X)} = \min\, \left\{X-m+1,N-2m+1 \right\}
    \frac{N!(N-2m+1)}{m!(N-m+1)!}. 
\end{equation}
The excitation number $X$ limits the dimension from above to $X-m+1$, resulting in the minimum between the two values. This dimensional structure is presented in Table~\ref{dimensiontableN9} for $N=9$. In addition to the above, we see the symmetric nature of the binomial coefficient in the total manifold dimension Eq.~\eqref{eqn:manifold_dimension}. This ``locking'' of the irrep dimensions will affect later results in Section~\ref{sect:eigenstates_selection_rules_and_emission}.

\begin{figure}[t]   
    \centering
    \includegraphics[width=.95\linewidth]{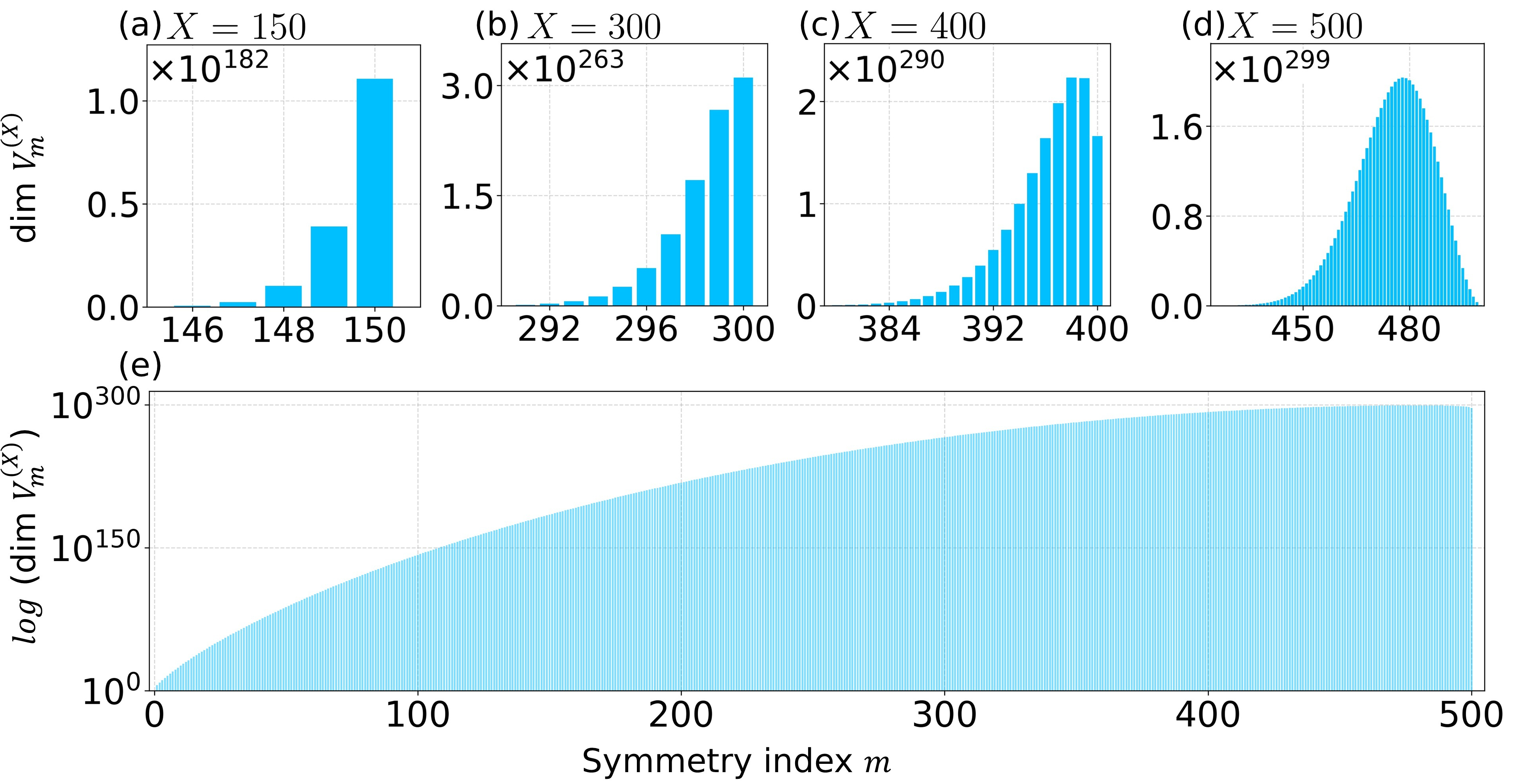}
    \caption{Dimensions of the irreps $V_m^{(X)}$ for $N=1000$ for $X \in \{150, 300, 400, 500\}$. In (a--b) the dark states dominate in multiplicity, while dark polaritons become progressively comparable with growing $X$. In (c) the dark polaritons overtake the dark states. In (d) the dark states disappear completely, and the dark polaritons end up in a smooth distribution peaked at high $m$. Panel (e) shows the logarithms of the $X=500$ dimensions, revealing the relative sizes invisible to the linear plots.}
    \label{fig:MultiplicityN10E3}
\end{figure}

Fig.~\ref{fig:MultiplicityN10E3} shows the dimensions given by Eq.~\eqref{eqn:dimension_formula} as functions of the symmetry index $m$ for $N=1000$ and $X \in \{150, 300, 400, 500\}$. Panels (a--d) show the high-$m$ regions in linear scale, while the entire range of $m$ for $X=500$ is considered in (e), with the dimension plotted in logarithmic scale.
We see that the irrep maximal in $m$ holds the highest multiplicity up to $X \approx N/3$, after which the dimensions approach a Poissonian-like distribution near the high-$m$ side of the distribution. We also see that the low-$m$ irreps become comparably insignificant in multiplicity, and that the multiplicity of the $m=X$ irrep approaches zero as $X \rightarrow \frac{N}{2}$.

The structure can be easily understood by comparing to the usual $X=1$ case. The representation $\mathcal{H}_m^{(1)}$ is two-dimensional with multiplicity $1$, and $\mathcal{H}_1^{(1)}$ is one-dimensional with multiplicity $N-1$. This is the well-known case of the upper polariton (UP), lower polariton (LP), and $N-1$ dark states~\cite{Embrace_the_darkness}. The $X=2$ case has also been previously studied~\cite{Siltanen_AOM,Campos-Gonzalez-Angulo_2}. As we will show in the next section, one can identify the subspace $V_0^{(X)}$ as \textit{multipolaritons}~\cite{Borges_Dark_Polariton} and $V_X^{(X)}$ as dark states ($X<N/2$). The representations in-between correspond to \textit{dark polaritons}, recognized by their photonic content being between these two extremal regimes~\cite{Campos-Gonzalez-Angulo_1,DelPo2020-yx}. Fig.~\ref{fig:MultiplicityN10E3} then shows that the dominant role of dark states is overtaken by dark polaritons and that the well-known problem of the large number of dark states becomes more subtle at higher excitation numbers. Thus, radiant processes from high-$m$ irreps might become statistically relevant~\cite{Sanchez-Barquilla,Mandal2023-jz}.

\begin{table}    
\begin{center}
\begin{tabular}{>{\columncolor{LightGoldenrod}}c|c|c|c|c|c|c} 
\rowcolor{LightSteelBlue} $X$ & $V_0^{(X)}$ & $V_1^{(X)}$ & $V_2^{(X)}$ & $V_3^{(X)}$ & $V_4^{(X)}$ &$\mathcal{H}^{(X)}$\\
\hline  
0 &\scriptsize 1   &\scriptsize               &\scriptsize                           & \scriptsize                              &\scriptsize                               & 1\\
1 &\scriptsize 2   &\scriptsize $1\cdot (N-1)$         &\scriptsize                           &\scriptsize                               &\scriptsize                               & 10\\
2 &\scriptsize 3   &\scriptsize $2\cdot(N-1)$ &\scriptsize $1\cdot \frac{1}{2}N(N-3)$       &\scriptsize                               &\scriptsize                               & 46\\
3 &\scriptsize 4   &\scriptsize $3\cdot(N-1)$ &\scriptsize $2\cdot \frac{1}{2}N(N-3)$&\scriptsize $1\cdot \frac{1}{6}N(N-1)(N-5)$      &\scriptsize                               & 130\\
4 &\scriptsize 5   &\scriptsize $4\cdot(N-1)$ &\scriptsize $3\cdot \frac{1}{2}N(N-3)$&\scriptsize $2\cdot\frac{1}{6}N(N-1)(N-5)$&\scriptsize $1\cdot \frac{1}{24}N(N-1)(N-2)(N-7)$& 256\\
5 &\scriptsize 6   &\scriptsize $5\cdot(N-1)$ &\scriptsize $4\cdot \frac{1}{2}N(N-3)$&\scriptsize $3\cdot\frac{1}{6}N(N-1)(N-5)$&\scriptsize $2\cdot\frac{1}{24}N(N-1)(N-2)(N-7)$& 382\\
6 &\scriptsize 7   &\scriptsize $6\cdot(N-1)$ &\scriptsize $5\cdot \frac{1}{2}N(N-3)$&\scriptsize $4\cdot\frac{1}{6}N(N-1)(N-5)$&\scriptsize $2\cdot\frac{1}{24}N(N-1)(N-2)(N-7)$& 466\\
7 &\scriptsize 8   &\scriptsize $7\cdot(N-1)$ &\scriptsize $6\cdot \frac{1}{2}N(N-3)$&\scriptsize $4\cdot\frac{1}{6}N(N-1)(N-5)$&\scriptsize $2\cdot\frac{1}{24}N(N-1)(N-2)(N-7)$& 502\\
8 &\scriptsize 9   &\scriptsize $8\cdot(N-1)$ &\scriptsize $6\cdot \frac{1}{2}N(N-3)$&\scriptsize $4\cdot\frac{1}{6}N(N-1)(N-5)$&\scriptsize $2\cdot\frac{1}{24}N(N-1)(N-2)(N-7)$& 511\\
9 &\scriptsize 10  &\scriptsize $8\cdot(N-1)$ &\scriptsize $6\cdot \frac{1}{2}N(N-3)$&\scriptsize $4\cdot\frac{1}{6}N(N-1)(N-5)$&\scriptsize $2\cdot\frac{1}{24}N(N-1)(N-2)(N-7)$& 512
\end{tabular}
\end{center}
\caption{Dimensions of the representation $V_m^{(X)}$ and the entire manifold $\mathcal{H}^{(X)} = \bigoplus_m V_m^{(X)}$, determined by $X$ and $m$ by Eq.~\eqref{eqn:dimension_formula}, for $N=9$. This value is substituted only to the total manifold dimension to clarify its limiting effect on the dimensions of the irreps $\mathcal{H}_m^{(X)}$.}
\label{dimensiontableN9}
\end{table}

\subsection{Example: The first two manifolds}

Let us present the diagonalization of the first two excitation manifolds to set what we seek to generalize. The first manifold Hamiltonian can be decomposed according to Eq.~\eqref{eqn:manifold_block_decomposition} into

\begin{align}   
    H^{(1)} &= H^{(1)}_0 \oplus (\identity_{N-1}\otimes H^{(1)}_1), \\
    H^{(1)}_0 &= \begin{pmatrix}
        E_{TLS} & \sqrt{N}g \\
        \sqrt{N}g & E_c
    \end{pmatrix},~~ H^{(1)}_1 = E_{TLS},
\end{align}
leading to the UP, LP, and $N-1$ dark states

\begin{align}   
    \ket{P_+} &= \frac{\alpha^{(1)}_0}{\sqrt{N}}  \sum_{n=1}^N \ket{e_n}\ket{0} 
    + \alpha^{(1)}_1 \ket{\mathcal{G}}\ket{1}, \\
    \ket{P_-} &= \frac{\alpha^{(1)}_1}{\sqrt{N}}  \sum_{n=1}^N \ket{e_n}\ket{0} 
    - \alpha^{(1)}_0 \ket{\mathcal{G}}\ket{1}, \\
    \ket{D_k} &= \frac{1}{\sqrt{N}} \sum_{n=1}^N e^{i2\pi nk/N} \ket{e_n}\ket{0},
    ~k=1,...,N-1,
\end{align}
with the (first-order) Hopfield coefficients and eigenvalues satisfying

\begin{align}   
    |\alpha^{(1)}_0|^2 &= \frac{1}{2}\left(1+ \frac{E_{TLS}-E_c}{\sqrt{(E_{TLS}-E_c)^2+4g_N^2}}
    \right), \label{eqn:first_ord_hopfield1}\\
    |\alpha^{(1)}_1|^2 &= \frac{1}{2}\left(1- \frac{E_{TLS}-E_c}{\sqrt{(E_{TLS}-E_c)^2+4g_N^2}}
    \right), \label{eqn:first_ord_hopfield2}\\
    E_\pm &= \frac{E_{TLS}+E_c}{2} \pm \sqrt{g^2_N+\frac{(E_{TLS}-E_c)^2}{4}}.\label{eqn:first_manifold_energies}
\end{align}
The dark states naturally carry the eigenvalue $E_{TLS}$ and we have denoted $g_N := \sqrt{N}g$. For the rest of this paper, we focus on the resonant case of $E_{TLS} = E_c$. With this zero detuning, the expressions in Eqs.~\eqref{eqn:first_ord_hopfield1}--\eqref{eqn:first_manifold_energies} simplify in a clear way. This shows us that the TLS-cavity energy difference acts as a small disturbance to a more balanced situation. We focus on the resonant case (unless otherwise specified), which allows for further analytic solvability, with this perturbative picture in mind.

Compared to the $X=1$ manifold, a new irrep becomes available for $X=2$, leading to the Hamiltonian decomposition~\cite{Campos-Gonzalez-Angulo_2}
\begin{equation}   
\begin{split}
    H^{(2)} = H_0^{(2)} \oplus &\left( \identity_{N-1} \otimes H_1^{(2)} \right)
    \oplus \left( \identity_{\frac{1}{2}N(N-3)} \otimes H_2^{(2)} \right) \\
    = \underbrace{\begin{psmallmatrix}
        2E_{TLS} & \sqrt{2}\sqrt{N-1}g & 0 \\
        \sqrt{2}\sqrt{N-1}g & 2E_{TLS} &\sqrt{2}\sqrt{N}g \\
        0 & \sqrt{2}\sqrt{N}g & 2E_{TLS}
    \end{psmallmatrix}}_{m=0}   &\oplus~ \underbrace{\identity_{N-1} \otimes
    \begin{psmallmatrix}
        2E_{TLS} & \sqrt{2}\sqrt{N-1}g \\
        \sqrt{2}\sqrt{N-1}g & 2E_{TLS}
    \end{psmallmatrix}}_{m=1} \oplus 
    \underbrace{2E_{TLS} \identity_{\frac{1}{2}N(N-3)}}_{m=2},
\end{split}
\end{equation}
with the values of $m$ corresponding to multipolaritons, dark polaritons, and dark states, respectively. These correspond to eigenstates of the form

\begin{align}   
    \ket{E^{(2,0)}_{i,0}} &=
    \alpha_0^{(2)}\sqrt{\frac{2}{N(N-1)}}\sum_{m<n}\ket{e_me_n}\ket{0} 
    + \alpha_1^{(2)}\frac{1}{\sqrt{N}}\sum_n \ket{e_n}\ket{1}
    + \alpha_2^{(2)}\ket{\mathcal{G}}\ket{2}, \\
    \ket{E^{(2,1)}_{0,k}} &=  \alpha_1^{(1)}\sum_{m<n} c_{mn}^{(k)} \ket{e_me_n}\ket{0}
    + \alpha_0^{(1)}\sum_n c_n^{(k)}\ket{e_n}\ket{1}, \\
    \ket{E^{(2,1)}_{1,k}} &=  \alpha_0^{(1)}\sum_{m<n} c_{mn}^{(k)} \ket{e_me_n}\ket{0}
    - \alpha_1^{(1)}\sum_n c_n^{(k)}\ket{e_n}\ket{1}, \\
    \ket{E^{(2,2)}_{0,(k,l)}} &= \sum_{m<n} c_{mn}^{(kl)}\ket{e_me_n}\ket{0},
\end{align}
where $\ket{E^{(X,m)}_{k,p}}$ refers to the $k$th eigenstate (in increasing energy) within the $X$th manifold and $m$th irrep, with $p$ indexing equivalent states of given multiplicity. These states correspond to the eigenenergies and (second-order) Hopfield coefficients

\begin{equation}   
\begin{split}
    E^{(2,0)}_i &\in \left\{2E_{TLS} - \sqrt{2}\sqrt{N-1}g,~ 2E_{TLS},~ 2E_{TLS}+\sqrt{2}\sqrt{N-1}g \right\}, \\
    E^{(2,1)}_i &\in \left\{2E_{TLS} - \sqrt{2N-2}g,~ 2E_{TLS}+\sqrt{2N-2}g \right\},~
    E^{(2,2)}_{0} = 2 E_{TLS}, \\
    \alpha_0^{(2)} &\in \left\{ \frac{1}{2},~ -\frac{1}{\sqrt{2}},~ \frac{1}{2} \right\}, ~
    \alpha_1^{(2)} \in \left\{ -\frac{1}{\sqrt{2}},~ 0,~ \frac{1}{\sqrt{2}} \right\}, ~
    \alpha_2^{(2)} \in \left\{ \frac{1}{2},~ \frac{1}{\sqrt{2}},~ \frac{1}{2} \right\},
\end{split}
\end{equation}
accordingly for $i=0,1,2$. Note that the $m=1$ eigenstates have the first-order Hopfield coefficients Eqs.~\eqref{eqn:first_ord_hopfield1}--\eqref{eqn:first_ord_hopfield2}.


\section{Irrep bases and eigenvalues}\label{sect:Irrep_bases_and_eigenvalues}

\subsection{Solving a reduced matrix form}

In what follows, we generalize the examples above. We find proper bases for all the irreps for given $N$ and $X$ to write the matrix elements of the Hamiltonian in symmetric form. We then use this form to derive a variety of properties for the systems considered by the model and to solve the eigensystem.

Take the unnormalized basis of symmetric states of the TC system~\cite{Dicke}. Add irrep information carrying complex-valued \textit{symmetric coefficients} $c_{\mathcal{I}}^{(\mathcal{J})}$, where $\mathcal{I}$ and $\mathcal{J}$ are index sets. These are determined by the action of the Young symmetrizer on the $N$-TLS states~\cite{Sagan,Stevens,Fulton1991RepresentationTA}. Let us denote arbitrary vector form states of $V_m^{(X)}$ in this basis as

\begin{align}   
    \Vec{\alpha}^{(X)}_m &:=(\alpha_0, \alpha_1, \dots, \alpha_{X-m})^{\text{T}},
\end{align}
where $\alpha_i$ is the amplitude of a symmetric state with $n$ photonic excitations, for example
\begin{align}
    \Vec{\alpha}_m^{(X)} &= 
    \alpha_0 \sum_{m<n}c_{mn}^{(k)}\ket{e_m e_n}\ket{0}
    + \alpha_1 \sum_n c_n^{(k)}\ket{e_n}\ket{1}\text{.}
\end{align} 
By orthogonality of the irreps, one finds that the symmetric coefficients obey

\begin{align}   
    \sum_{\mathcal{I}} c_{\mathcal{I}}^{(\mathcal{J})} c_{\mathcal{I}}^{(\mathcal{J}')*} &= \prod_{\mathcal{J,J'}} \delta_{jj'}, ~~~~|\mathcal{J}|=|\mathcal{J'}|, \\
    \sum_{\mathcal{I}} c_{\mathcal{I}}^{(\mathcal{J})} c_{\mathcal{I}}^{(\mathcal{J}')*} &= 0, ~~~~ |\mathcal{J}|\not=|\mathcal{J'}|,\\
    \sum_{\mathcal{I}} c_{\mathcal{I}}^{(\mathcal{J})} &=0, 
\end{align}
where $j \in \mathcal{J}, j' \in \mathcal{J}'$, and $\mathcal{I},\mathcal{J}$, and $\mathcal{J}'$ are index sets of finite order, with $|\mathcal{J}| = m$ and $|\mathcal{I}| = X_{\y{TLS}}$, the latter being the number of atomic excitations present. We further explicitly define $c_{\mathcal{I}}^{(\emptyset)}:=1$ for completeness. It should be noted that it is justified to take this basis into use, as these are not yet physical states. The key point is that we can use these as ansatz to find the true normalized bases through similarity transformations.

The interaction Hamiltonian only allows transitions between basis states of adjacent photonic content, and so Eqs.~\eqref{eqn:free_Hamiltonian}--\eqref{eqn:interaction_Hamiltonian} expanded in the symmetric basis is a tridiagonal matrix. Under the resonance condition $E_{TLS}=E_c$, the free Hamiltonian is a scalar matrix $XE_{TLS} \identity$, and so the eigenvalues and eigenvectors of the system depend only on those of the rescaled interaction matrix
\begin{align}\label{eqn:A_matrix_form} 
    A &:= \frac{1}{g}(H_{TC}-E_{TLS} \identity) \notag \\
    &= \begin{psmallmatrix}
    0 & a_1'  &  &  &  &  &  &  \\
    a_1 & 0 & a_2'  &  &  &  &  &  \\
     & a_2 &0 & a_3'  &  &  &  &  \\
     &  & a_3 & 0 & \ddots &  &  &  \\
     &  &  & \ddots & \ddots & a_{X-m-2}' &  &  \\
     &  &  &  & a_{X-m-2} & 0 & a_{X-m-1}' &  \\
     &  &  &  &  & a_{X-m-1} & 0 & a_{X-m}' \\
     &  &  &  &  &  & a_{X-m} & 0
    \end{psmallmatrix}_{X-m+1},
\end{align}
where $(H_I)_{X-m+1} = gA$. If the original eigenvalue equation of the TC Hamiltonian reads $H_{\text{TC}} \ket{V^{(X)}_m}=E \ket{V^{(X)}_m}$, we have reduced the problem to solving the eigenvalues of $A$, $A \Vec{\alpha}=\lambda \Vec{\alpha}$, where $\lambda:= \frac{E-XE_{TLS}}{g}$.

Let us then find the matrix elements of $A$. Consider the two terms of the interaction Hamiltonian, $H_I^{[1]}:=\sum_{n=1}^N \ket{e_n}\bra{g_n}\hat{a}$ and $H_I^{[2]}:=\sum_{n=1}^N \ket{g_n}\bra{e_n}\hat{a}^{\dagger}$, acting on bare states of the type $\ket{V}=\ket{e^k}\ket{l}$, where $\ket{e^k}=\ket{e_{n_1} e_{n_2}\cdots e_{n_k}}$ , and  $k+l=X$. The cavity operators $\hat{a}~(\hat{a}^{\dagger})$ operating on the bare states simply correspond to factors of $\sqrt{l}\left(\sqrt{l+1}\right)$. For the $H_I^{[1]}$ input state $\ket{e^{k-1}}$ and output state $\ket{e^k}$, $\binom{k}{k-1} =k$ states can transform into $\ket{e^k}$, so $\sum_{n=1}^N \ket{e_n}\bra{g_n}$ leaves a factor of $k$. Similarly for $H_I^{[2]}$ input state $\ket{e^{k+1}}$ and output state $\ket{e^{k}}$, $N-k$ of states $\ket{e^{k+1}}$ can transform into $\ket{e^{k}}$, leaving a factor of $N-k$. The symmetric coefficients change according to the corresponding irrep. We summarize these results in Table~\ref{BareOperation}.

The procedure we want to generalize is as follows: \textit{Find the matrix elements in the symmetric basis and transform them into the properly normalized basis, for general $N$, $X$, and $m$}. If we apply the results of Table~\ref{BareOperation} on the symmetric basis and note that the off-diagonal sequences contain $X-m$ terms, we find that the matrix elements are $a_l^{(C)}\cdot a_l^{\downarrow}$ and $a_l^{(C)}\cdot a_l^{\uparrow}$ for the sub- and superdiagonal sequences with values given in Table~\ref{BareElements1}. 

\begin{table}[t] 
\begin{center}
\begin{tabular}{c|c|c} 
\rowcolor{LightSteelBlue} Operator  & Input state & Output state\\
\hline 
$\sum_{n=1}^N \ket{e_n}\bra{g_n}\hat{a}$ & $\ket{e^k}\ket{X-k}$ 
& $(k+1)\sqrt{X-k}\ket{e^{k+1}}\ket{X-k-1}$  \\
$\sum_{n=1}^N \ket{g_n}\bra{e_n}\hat{a}^{\dagger}$ & $\ket{e^k}\ket{X-k}$ 
& $(N-(k-1))\sqrt{X-k+1}\ket{e^{k-1}}\ket{X-k+1}$  
\end{tabular}
\end{center}
\caption{The TC interaction Hamiltonian operating on bare states.}
\label{BareOperation}
\end{table}

\begin{table}[t]
\begin{center}
\begin{tabular}{c|c|c} 
\rowcolor{LightSteelBlue} Shared cavity term  $a_l^{(C)}$& Subdiagonal term $a_l^{\downarrow}$ & Superdiagonal term $a_l^{\uparrow}$\\
\hline 
$\sqrt{1}$ &   $N-(X-1)$ & $X$       \\
$\sqrt{2}$ &   $N-(X-2)$ & $X-1$     \\
$\sqrt{3}$ &   $N-(X-3)$ & $X-2$     \\
$\vdots$  &   $\vdots$  & $\vdots$  \\
$\sqrt{X-(m-1)}$ &   $N-(m-1)$ & $m+2$     \\
$\sqrt{X-m}$ &   $N-m$     & $m+1$
\end{tabular}
\end{center}
\caption{Cofactors making up the sequences of matrix elements along the sub- and superdiagonals of the symmetric basis Hamiltonian, with $X-m$ terms along each sequence.}
\label{BareElements1}
\end{table}

The Hamiltonian is diagonalizable and Hermitian and therefore symmetric and real in some basis [$a_l = a_l' \in \mathbb{R}$ in Eq.~\eqref{eqn:A_matrix_form}]. This coincides with the properly normalized basis, and the proper matrix $A$ can be found recursively: The matrix $P$ carrying the basis transformation $PAP^{-1}$ is diagonal, and the first element is $P_{00}=1$. The second element is $P_{11}=\frac{\sqrt{X}}{\sqrt{N-X+1}}$, comprising of the square roots of the first terms in the sub- and superdiagonal sequences. The rest are then determined recursively as $P_{ll} = P_{l-1,l-1}\cdot \frac{\sqrt{c^{\uparrow}_l}}{\sqrt{c^{\downarrow}_l}}$. The new matrix element coefficients are of the form $c_l := c_l^{(N)}\cdot c_l^{(C)} \cdot c_l^{(M)}$, where the cofactors $c_l^{(i)}$ represent normalization ($N$), cavity ($C$), and TLS (matter, $M$) contributions. The matrix elements $(A)_{ij} = c_l \delta^{|i-j|}_1 \delta^{\min{\{i,j\}}}_l$ are presented in Table~\ref{BareElements2}.

\begin{table}[t] 
\begin{center}
\begin{tabular}{c|c|c} 
\rowcolor{LightSteelBlue} Cavity term  $c_l^{(C)}$ & TLS term $c_l^{(M)}$ & Normalization term $c_l^{(N)}$ \\
\hline 
$\sqrt{1}$ &   $\sqrt{X}$ & $\sqrt{N-(X-1)}$       \\
$\sqrt{2}$ &   $\sqrt{X-1}$ & $\sqrt{N-(X-2)}$     \\
$\sqrt{3}$ &   $\sqrt{X-2}$ & $\sqrt{N-(X-3)}$     \\
$\vdots$  &   $\vdots$  & $\vdots$  \\
$\sqrt{X-(m-1)}$ &   $\sqrt{m+2}$ & $\sqrt{N-(m+1)}$     \\
$\sqrt{X-m}$ &   $\sqrt{m+1}$     & $\sqrt{N-m}$
\end{tabular}
\end{center}
\caption{Cofactors making up the sequences of matrix elements $c_l = c_l^{(N)}\cdot c_l^{(C)} \cdot c_l^{(M)} = \sqrt{l(X+1-l)(N-X+l)}$ along the first off-diagonals of the proper basis Hamiltonian $H_I$, with $X-m$ terms in each sequence.}
\label{BareElements2}
\end{table}

As a real symmetric tridiagonal matrix, $A$ has many useful properties. An immediate consequence is that all the eigenvalues are simple and real. This means that the representation-theoretic multiplicity then determines also the multiplicities of the eigenvalues. If $D := \y{diag}(\dots,-1,1,-1,1)$, it is easy to see that $A$ and $D$ anticommute, so the spectrum of $A$ is symmetric about zero (see Appendix \ref{app:Symmetric_spectrum}). A direct corollary is that $0$ is an eigenvalue for $X-m \equiv 0 \mod{2}$, and that one has to solve only one half of the spectrum, allowing for significant numerical optimization. As $m$ restricts only the length of the matrix element sequence above---within a given excitation manifold---the matrix corresponding to $m'=m-1$ is a principal submatrix of that of $m$. It follows from the Cauchy interlacing theorem that the eigenvalues of $H^{(X)}_m$ and $H^{(X)}_{m-1}$ alternate~\cite{Cauchy_interlacing}. Starting recursively from $H^{(X)}_X$, it follows that $H^{(X)}_0$ has the extremal eigenvalues of an excitation manifold. These extremal values have a bound proportional to the matrix elements according to the Geršgorin disc theorem~\cite{Horn_Johnson_1985}. By estimating the values of Table~\ref{BareElements2} upward for $m=0$ by $\sqrt{l\cdot(X-l)} \leq X/2$ and $c^{(N)}_l \leq \sqrt{N}$, a Geršgorin bound of $X\sqrt{N}$ is found. Later calculations show that this is very close to the true extremal values.

With the bases and matrix elements derived above, we are able to calculate the eigenvalues $E^{(X,m)}_k = XE_{TLS} +\lambda^{(X,m)}_k g$, the eigenvectors $\ket{E^{(X,m)}_{k,p}}$, their probability amplitudes squared $|\alpha_i|^2$, average photonic content $\langle n_c \rangle$, and average TLS excitation content $\langle n_{\y{TLS}} \rangle$. These quantities are presented in Table~\ref{Eigensystems1} for $X=1,2,3$, where we omit writing the indices $(X,m)$ where it does not cause confusion. The vector components $\alpha_i$ have convergent $N$-dependence, strong already at realistic values of $N>10^6$, so we have calculated $\lim_{N\rightarrow \infty} \alpha_i$ for the tabled values (more on the limit in Section~\ref{sect:eigenstates_selection_rules_and_emission}). We will continue by inspecting each quantity at a time.

\subsection{Matrix spectra}

\begin{figure}   
    \centering
    \includegraphics[width=1\linewidth]{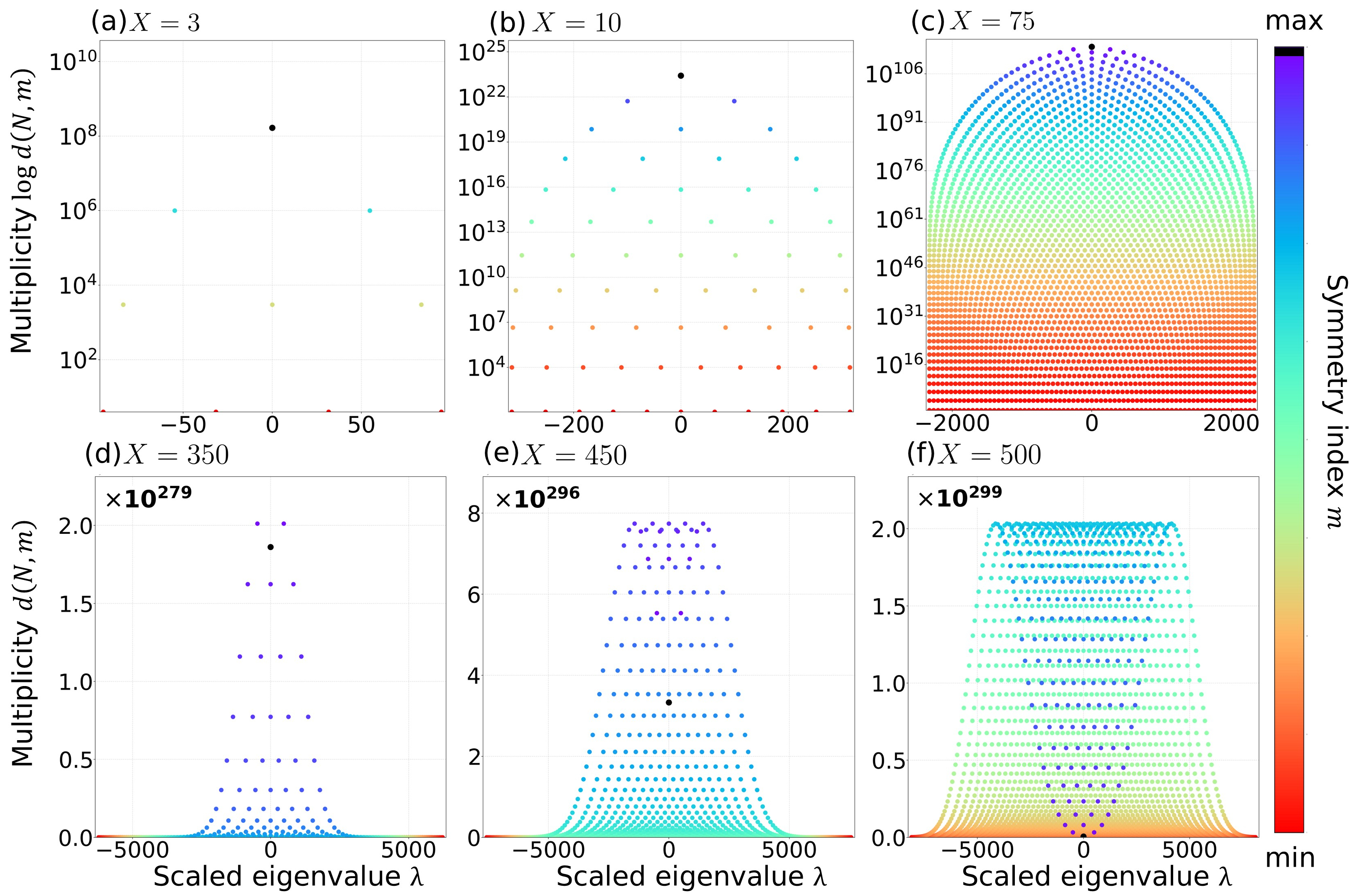}
    \caption{Eigenvalue distributions of $A$ for $N=1000$ and $X\in\{3,10,75,350,450,500\}$. Panels (a--c) are in logarithmic scale and $0\leq m\leq X$, showing a trade-off between the multiplicity and energy density for high and low $m$. Panels (d--f) show linearly for $0.8X\leq m\leq X$, how the statistically most relevant states compare at high excitation numbers. The distributions represent a unitless, centered-around-zero energy spectrum for the system, scalable to arbitrary $E_{TLS}=E_c$ and $g$. The TLS excitation content of the states is proportional to $m$, such that red points have higher photonic content, and black points refer to dark states. The panels then show a statistical comparison between the variety of states with differing photonic contents.}
    \label{fig:spectralweight}
\end{figure}

We see from Table~\ref{Eigensystems1} that the eigenvalues tend to follow a pattern of nested square roots containing polynomials of $N$. While it may be possible, an analytical expression for general $N$, $X$, and $m$ seems to be highly non-trivial to find. The properties of $A$ described above allow for heavy optimization with numerical methods. For the following calculations, we used the NumPy library for Python. Fig.~\ref{fig:spectralweight} shows the eigenvalues $\lambda$ of $A$ against their multiplicities for $N=1000$ and $X\in\{3,10,75,350,450,500\}$. The color encodes the symmetry index $m$. Panels (a--c) are in logarithmic scale and $0\leq m\leq X$. Panels (d--f) are in linear scale and $0.8X\leq m\leq X$. Recall that $m=0$ corresponds to multipolaritons and $m=X$ to dark states, with maximal and minimal $\langle n_c \rangle$, respectively. 

We see that the statistical relevance of differing energies centers near the dark states. This effect is countered by the fact that lower-$m$ irreps are more dense in energy states due to higher dimensions and the spectrum being bounded. The eigenvalues are mirror-symmetric around zero. We also recognize the effects on the relevance of dark states seen already in Fig.~\ref{fig:MultiplicityN10E3}. Furthermore, photonic energy states appear at the dark-state energy for even $X-m$ due to the symmetric spectrum, although this is perturbed by the detuning, i.e., $E_{TLS} \not= E_c$~\cite{Siltanen_AOM}.

Note that the multiplicities grow considerably between panels. This growth is stunted after $X>\frac{N}{2}$ as no new irreps appear and the energy levels only get denser (Table~\ref{dimensiontableN9}). This can be understood as approaching a continuous limit for the energy spectrum. As the $S_N$ irreps define the TLS excitation content of the eigenstates, dark states must disappear after this limit, with new excitations going to the cavity.

\begin{table}[ht] 
\begin{center}
\begin{tabular}{>{\columncolor{LightGoldenrod}}c|c|c|c|c}
\rowcolor{LightSteelBlue} \small$\ket{E_k} $ &\small $\ket{E_k}=(\alpha_i)_{i=0}^{X-m}$
&\small $\left\{ |\alpha_i|^2 \right\}$ &\small $\langle n_c \rangle$ &\small $\langle n_{TLS} \rangle$\\
\hline 
$V_0^{(1)}$ & \multicolumn{4}{>{\columncolor{MistyRose}}c}{$\lambda_k \in \left\{ \pm \sqrt{N} \right\}$} \\
\hline
$\ket{E_0}$ &  $\frac{1}{\sqrt{2}}\left( -1,1\right)^{\mathrm{T}}$  
& $\left\{ \frac{1}{2}, \frac{1}{2} \right\}$ & $1/2$ & $1/2$    \\
$\ket{E_1}$ & $\frac{1}{\sqrt{2}}\left( 1,1\right)^{\mathrm{T}}$
& $\left\{ \frac{1}{2}, \frac{1}{2} \right\}$ & $1/2$ & $1/2$ \\
\hline
$V_0^{(2)}$ & \multicolumn{4}{>{\columncolor{MistyRose}}c}{$\lambda_k \in \left\{ 0, \pm \sqrt{4N-2} \right\}$}   \\
\hline
$\ket{E_0}$ & $\frac{1}{2} \left( 1,-\sqrt{2},1 \right)^{\mathrm{T}}$  
& $\left\{ \frac{1}{4}, \frac{1}{2} ,\frac{1}{4} \right\}$ & 1 & 1 \\
$\ket{E_1}$ & $\frac{1}{\sqrt{2}} \left( -1,0,1 \right)^{\mathrm{T}}$  
& $\left\{ \frac{1}{2},0 ,\frac{1}{2} \right\}$  & 1 & 1 \\
$\ket{E_2}$ & $\frac{1}{2} \left( 1,\sqrt{2},1 \right)^{\mathrm{T}}$  
& $\left\{ \frac{1}{4}, \frac{1}{2} ,\frac{1}{4} \right\}$ & 1 & 1 \\
\hline
$V_1^{(2)}$ & \multicolumn{4}{>{\columncolor{MistyRose}}c}{$\lambda_k \in \left\{\pm \sqrt{2N-2} \right\}$}   \\
\hline
$\ket{E_0}$ &  $\frac{1}{\sqrt{2}}\left( -1,1\right)^{\mathrm{T}}$  
& $\left\{ \frac{1}{2}, \frac{1}{2} \right\}$ & $1/2$ & $3/2$    \\
$\ket{E_1}$ & $\frac{1}{\sqrt{2}}\left( 1,1\right)^{\mathrm{T}}$
& $\left\{ \frac{1}{2}, \frac{1}{2} \right\}$ & $1/2$ & $3/2$ \\
\hline
$V_0^{(3)}$ & \multicolumn{4}{>{\columncolor{MistyRose}}c}{\footnotesize $\lambda_k \in \left\{ \pm \sqrt{5N-5+\sqrt{16N^2 -32N+25}}, 
~ \pm  \sqrt{5N-5-\sqrt{16N^2-32N+25}} \right\}$}   \\
\hline
$\ket{E_0}$ & $\frac{1}{2\sqrt{2}} \left( 
-1, \sqrt{3}, -\sqrt{3} , 1 \right)^{\text{T}}$ & $\left\{ \frac{1}{8}, \frac{3}{8}, \frac{3}{8}, \frac{1}{8} \right\}$ & $3/2$ & $3/2$ \\
$\ket{E_1}$ & $\frac{\sqrt{3}}{2\sqrt{2}} \left( 
1, -\frac{1}{\sqrt{3}} , -\frac{1}{\sqrt{3}} , 1 \right)^{\text{T}}$ & $\left\{ \frac{3}{8}, \frac{1}{8}, \frac{1}{8}, \frac{3}{8} \right\}$ & $3/2$ & $3/2$ \\
$\ket{E_2}$ & $\frac{\sqrt{3}}{2\sqrt{2}} \left( 
-1, -\frac{1}{\sqrt{3}} ,\frac{1}{\sqrt{3}} , 1 \right)^{\text{T}}$ & $\left\{ \frac{3}{8}, \frac{1}{8}, \frac{1}{8}, \frac{3}{8} \right\}$ & $3/2$ & $3/2$ \\
$\ket{E_3}$ & $\frac{1}{2\sqrt{2}} \left( 
1, \sqrt{3},\sqrt{3} , 1 \right)^{\text{T}}$ & $\left\{ \frac{1}{8}, \frac{3}{8}, \frac{3}{8}, \frac{1}{8} \right\}$ & $3/2$ & $3/2$ \\
\hline
$V_1^{(3)}$ & \multicolumn{4}{>{\columncolor{MistyRose}}c}{$\lambda_k \in \left\{ 0, \pm \sqrt{7N-10} \right\}$}   \\
\hline
$\ket{E_0}$ & $\frac{\sqrt{2}}{\sqrt{7}} \left( \frac{\sqrt{3}}{2},-\frac{\sqrt{7}}{2},1 \right)^{\mathrm{T}}$  
& $\left\{ \frac{3}{14}, \frac{7}{14} ,\frac{4}{14} \right\}$ & $1+1/14$ & $2-1/14$ \\
$\ket{E_1}$ & $\frac{\sqrt{3}}{\sqrt{7}} \left( -\frac{2}{\sqrt{3}},0,1 \right)^{\mathrm{T}}$  
& $\left\{ \frac{4}{7},0 ,\frac{3}{7} \right\}$  & $1-2/14$ & $2+2/14$ \\
$\ket{E_2}$ & $\frac{\sqrt{2}}{\sqrt{7}} \left( \frac{\sqrt{3}}{2},\frac{\sqrt{7}}{2},1 \right)^{\mathrm{T}}$  
& $\left\{ \frac{3}{14}, \frac{7}{14} ,\frac{4}{14} \right\}$ & $1+1/14$ & $2-2/14$ \\
\hline
$V_2^{(3)}$ & \multicolumn{4}{>{\columncolor{MistyRose}}c}{$\lambda_k \in \left\{ \pm \sqrt{3N-6} \right\}$}   \\
\hline
$\ket{E_0}$ &  $\frac{1}{\sqrt{2}}\left( -1,1\right)^{\mathrm{T}}$  
& $\left\{ \frac{1}{2}, \frac{1}{2} \right\}$ & $1/2$ & $5/2$    \\
$\ket{E_1}$ & $\frac{1}{\sqrt{2}}\left( 1,1\right)^{\mathrm{T}}$
& $\left\{ \frac{1}{2}, \frac{1}{2} \right\}$ & $1/2$ & $5/2$ \\
\hline
$V_X^{(X)}$ & $\ket{E^{(X,m)}_0} = (1)$,  $\lambda = 0$  & $\left\{ 1 \right\}$ & $0$ & $X$  
\end{tabular}
\end{center}
\caption{Eigensystems of $A$ for $X=1,2,3$. Presented in order are kets, vector components with $\lim_{N \rightarrow \infty}$, probability amplitudes squared, average photonic content, and average TLS content. The irreps are divided by listing the subspace $V_m^{(X)}$ and the scaled eigenvalues $\lambda_k$. Energies $E_k=XE_{TLS}+\lambda_kg$ are in increasing order of $k$.}
\label{Eigensystems1}
\end{table}

\clearpage

The center of the \textit{true} eigenvalues is $XE_{TLS}$, and so the difference between manifold centers is $E_{TLS}$. However, the bounds of the spectra grow proportional to $X\sqrt{N}$, and numerical calculations suggest that the latter factor is in the order of $\sqrt{N-\mathcal{O}(X)}$. This means that in the full energy spectrum of the TC model, the excitation manifolds start to overlap. For the approximate values of $E_{TLS} = 3.0~\y{eV}$ and $g=0.3~\y{meV}$~\cite{Abdelmagid_Gruyter}, we can estimate the overlap by
\begin{equation}   
\begin{split}
    E_{min}^{(X)}  &<  E_{max}^{(X-1)}\\
    \Leftrightarrow E_{TLS} &< (2X-1)\sqrt{N} g,
\end{split}
\end{equation}
giving the following values of $X$ for fixed $N$,
\begin{align}
    N=10^3&:~X>159, \\
    N=10^6&:~X>6,
\end{align}
understandable through the increasing Rabi split with growing $N$. The spectra start to overlap already at low excitation numbers. This shows that the picture of separate energy landscapes between manifolds (see, e.g., Ref.~\cite{Siltanen_AOM}) does not hold for realistic values of $N$ and $X$. 

We see from Table~\ref{Eigensystems1} that the photonic and TLS excitation contents follow approximately the patterns
\begin{align}\label{EQ-AVERAGE PHOTONIC CONTENT}   
    \langle n_c \rangle &\simeq \frac{X-m}{2}, \\
    \langle n_{TLS} \rangle &\simeq \frac{X+m}{2}\text{.}\label{EQ-AVERAGE TLS CONTENT}
\end{align}
The averages of the contents over the irreps $V^{(X)}_m$ are basis-independent and can be calculated in the symmetric basis to be exactly Eqs.~\eqref{EQ-AVERAGE PHOTONIC CONTENT}--\eqref{EQ-AVERAGE TLS CONTENT} (see Appendix~\ref{app:Average_excitation_contents}). This suggests that states with higher $m$ contribute more to material processes, such as singlet-singlet annihilation of excitons in organic molecules~\cite{Ruseckas_SSA}. Fig.~\ref{fig:spectralweight} also suggests that these processes are further amplified by large multiplicities, and that the contributing states lie close to the middle of the spectrum in energy.


\section{Eigenstates, selection rules, and emission energies}\label{sect:eigenstates_selection_rules_and_emission}

\subsection{Asymptotic eigensystem}

The true eigenvalues and eigenstates of the system can be calculated numerically with the results of the previous section. However, as analytical expressions are difficult to formulate, we will use the asymptotic limit $N \rightarrow \infty$ to find useful properties, which can then be shown to hold also without the limit. The limit has also been used as a starting point for a perturbative approach to find the true eigenstates for a given $N$~\cite{Sanchez_CUT-E}. This limit is reasonable, as the diverging $\sqrt{N}$ is balanced by the coupling strength $g$ decreasing with growing $N$.

Noting that for $n \in \mathbb{N}$, $\lim_{N\rightarrow\infty} \sqrt{N-n} = \lim_{N\rightarrow\infty} \sqrt{N}$, it can be seen that
\begin{align}   
    \lim_{N\rightarrow \infty}A &= \lim_{N\rightarrow \infty} \sqrt{N} \Tilde{A~} \notag \\
    \Leftrightarrow 0 &= \lim_{N \rightarrow \infty} 
    \left( \frac{1}{\sqrt{N}}A - \Tilde{A~}\right), \notag \\
    \Tilde{A~} &= 
    \begin{psmallmatrix}
        0 & \sqrt{X} & & & \\
        \sqrt{X} & 0 & \sqrt{2}\sqrt{X-1} & &  \\
         & \sqrt{2}\sqrt{X-1} & 0 & \ddots&  \\
         &  & \ddots & \ddots & \sqrt{X-m}\sqrt{m+1}\\
         & & & \sqrt{X-m}\sqrt{m+1} & 0
    \end{psmallmatrix},
\end{align}
and so $A$ and $\Tilde{A~}$ are simultaneously diagonalizable, meaning that they have the same eigenstates at high $N$. These are the eigenstates in Table~\ref{Eigensystems1}. The new matrix $\Tilde{A~}$ is much easier to solve, and it carries additional properties. For $m=0$ it is persymmetric, i.e., symmetric with respect to the anti-diagonal. Since $\Tilde{A~}$ is also symmetric, this is quantified as commuting with the exchange matrix $J$. This commutation makes the eigenstates of $\Tilde{A~}$ those of $J$ also, with eigenvalues $\pm 1$, giving $J$ the interpretation of a parity operator, where each eigenstate is labeled by an integer $\pm 1$
\begin{align}\label{eqn:Multipolariton_Parity}   
    \begin{cases}
        J \Vec{\alpha}^{(X)}_0 = \Vec{\alpha}^{(X)}_0, \\
        J \Vec{\alpha}^{(X)}_0 = -\Vec{\alpha}^{(X)}_0.
    \end{cases}
\end{align}
Component-wise, this means that the eigenstate components are (anti-)symmetric about the middle component depending on the parity $\pm1$.
There are equally many (or off by one for odd dimension) of the two parities, and in Appendix~\ref{app:Parity_alternation} we show that they must alternate with increasing eigenvalue.
We also notice that the $J$ and $D$ from Section~\ref{sect:Irrep_bases_and_eigenvalues} (anti-)commute for (even) odd dimensions, meaning that parity is (anti-)symmetric along the spectrum.
The converging factors of the true eigenstate components are square roots of positive rational functions of $N$, meaning that the strict parity of the high $N$ limit still shows as generalized parity in the signs of the true components. The binary ``sign pattern'' of an eigenstate then becomes a descriptive property of the said state.

\subsection{Allowed trajectories}

Let us then consider coupling with an environment of bosonic modes, with interactions given by the system part of a product interaction Hamiltonian~\cite{Breuer_Petruccione,Scala_Piilo}
\begin{align}\label{eqn:Environment_Interaction_Hamiltonian}
    H_{I,S} &= \hat{a}+\hat{a}^{\dagger}.
\end{align}
These operators conserve the cooperation number $S$, and so $m$ is also conserved (same applies for $\hat{S}$ and $\hat{S}^{\dagger}$). This means that the dynamics generated by Eq.~\eqref{eqn:Environment_Interaction_Hamiltonian} must be restricted within subspaces of the same irrep. For example, subsequent instances of emission ($\hat{a}$) will drive all the dark polaritons to dark states. This can be disturbed by symmetry-breaking processes, such as dephasing generated by, e.g., $\hat{S}_x$. 

Polaritonic systems often undergo internal processes at faster rates compared to environ\-ment-induced ones such as emission or non-radiative relaxation~\cite{Siltanen_MH}. Since a system naturally minimizes its energy, the environmental processes mostly concern minimal eigenstates of given irreps. This can be further restricted to the lowest-energy eigenstates of the totally symmetric irrep, if we allow dephasing. This allows us to focus on these states and find selection rules for the trajectories of the system.

One can show (Appendix \ref{app:Extremal_eigenstates}) that the eigenstates of $\Tilde{A~}$ with eigenvalues $X$ are given by the components
\begin{align}\label{eqn:Extremal_Eigenstate_Components}
    \ket{E^{(X,0)}_X}_l = \sqrt{\frac{1}{2^X}\binom{X}{l}},
\end{align}
for $0\leq l\leq X$. On the other hand, Geršgorin discs give an upper bound of $X$ for the eigenvalues, meaning that this is the maximal eigenstate. Because the spectrum is symmetric, the minimum energy is $-X$ with the eigenstate $D\ket{E^{(X,0)}_X} = \ket{E^{(X,0)}_0} \equiv \ket{E^{(X)}_{LMP}}$, i.e., the lowest multipolariton (LMP). A direct calculation in Appendix~\ref{app:Transition_probabilities} shows that these states have excitation contents $\frac{X}{2}$, equaling to the matrix element contributing to the radiative transition probability
\begin{align}\label{eqn:Transition_Probability}
    |\bra{E^{(X-1)}_{LMP}}\hat{a}\ket{E^{(X)}_{LMP}}|^2
    &= \bra{E^{(X)}_{LMP}}\hat{n}\ket{E^{(X)}_{LMP}}
    =\frac{X}{2}.
\end{align}

It can be further seen that the span of lowest-energy eigenstates is closed under $\hat{a}$. This means that the transition probability to other states must be equal to zero. 
We can evaluate the inner product with the Cauchy-Schwartz (CS) inequality
\begin{align}\label{eqn:Annihilation_Projection}
    \sqrt{\frac{X}{2}} = \bra{E^{(X-1)}_{LMP}}\hat{a}\ket{E^{(X)}_{LMP}}
    &\leq \sqrt{\braket{E^{(X-1)}_{LMP}}{E^{(X-1)}_{LMP}}\bra{E^{(X)}_{LMP}}\hat{a}^{\dagger}
    \hat{a}\ket{E^{(X)}_{LMP}}} \notag \\
    &= \sqrt{1\cdot \langle n_c \rangle ^{(X,0)}_0}
    = \sqrt{\frac{X}{2}},
\end{align}
where equality holds if and only if  $\hat{a}\ket{E^{(X)}_{LMP}} \in  \y{span}\left\{\ket{E^{(X-1)}_{LMP}}\right\}$, showing the claim by orthogonality of the eigenstates. By taking Hermitian conjugates of the above equations, one reaches the same result for $\hat{a}^{\dagger}$,
\begin{align}   
    \hat{a}  &:~V_0^{(X)} \rightarrow V_0^{(X-1)},~ \label{eqn:emission_channel}
    \ket{E^{(X)}_{LMP}} \mapsto \sqrt{\frac{X}{2}}\ket{E^{(X-1)}_{LMP}}, \\
    \hat{a}^{\dagger} &:~V_0^{(X)} \rightarrow V_0^{(X+1)},~
    \ket{E^{(X)}_{LMP}} \mapsto \sqrt{\frac{X+1}{2}}\ket{E^{(X+1)}_{LMP}}.
\end{align}

Next, we will consider three types of processes: emission $\mathcal{E}$, symmetry-preserving (or $m$-conserving) processes $\Phi$, and symmetry-breaking processes $\Psi$. If we decompose a general density operator under the irrep structure, $\rho = \bigoplus_m \rho_m$, these quantum channels can be written as follows.
\begin{align} \label{eqn:quantum_channels}
    \Phi:\mathcal{S}\left(\mathcal{H}^{(X)}\right) \rightarrow \mathcal{S}\left(\mathcal{H}^{(X)}\right),
    &~~~\Phi(\rho) = \bigoplus_m \Phi_m(\rho_m),  \\
    \Psi:\mathcal{S}\left(\mathcal{H}^{(X)}\right) \rightarrow \mathcal{S}\left(\mathcal{H}^{(X)}\right),
    &~~~\Psi(\rho) = \rho' \not= \bigoplus_m \Psi_m(\rho_m), 
\end{align}
where $\Phi_m$ and $\Psi_m$ are restrictions of $\Phi$ and $\Psi$ to $V_m^{(X)}$, and $\mathcal{S}(\mathcal{H}^{(X)})$ is the state space corresponding to the excitation manifold $\mathcal{H}^{(X)}$. The emission channel $\mathcal{E}$ is given by Eq.~\eqref{eqn:emission_channel}. An example of $\Phi$ could be Lindbladians generated by $\hat{a}$ or $\hat{S}$, and an example of $\Psi$ could be total TLS dephasing generated by the Pauli operator $\hat{S}_x$.

\begin{figure}[t]
    \centering
    \includegraphics[width=0.6\linewidth]{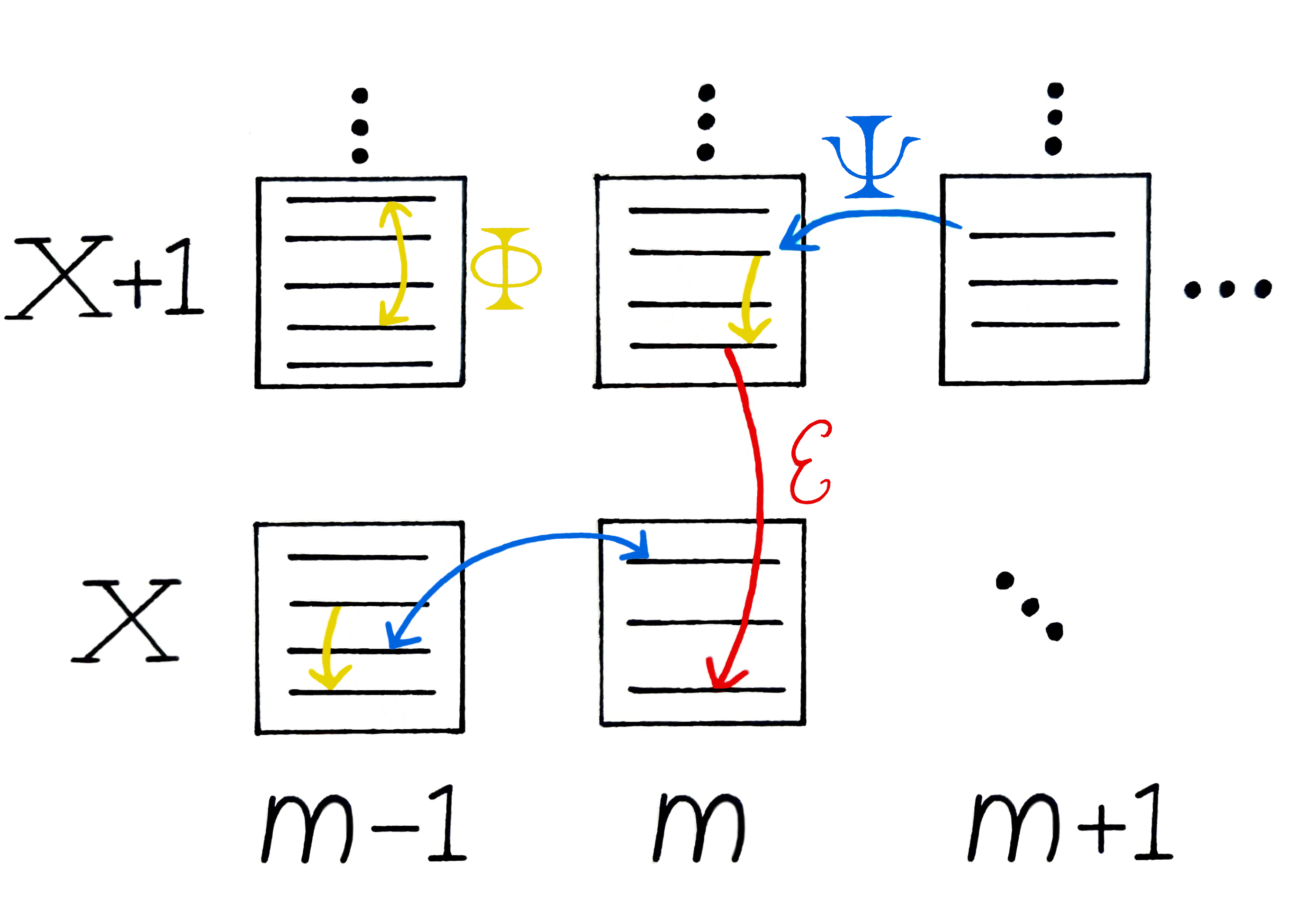}
    \caption{A schematic illustration of the quantum channels $\mathcal{E}$, $\Psi$, and $\Phi$. The boxes correspond to the representations $V_m^{(X)}$ and the lines inside to the energy eigenstates $\ket{E^{(X,m)}_k}$. $\mathcal{E}$ induces transitions between LMP states of adjacent excitation manifolds, $\Psi$ transitions between irreps within a given manifold, and $\Phi$ transitions between states within a given manifold and irrep.}
    \label{fig:quantum_channels}
\end{figure}

A physically allowed channel $\varphi$ has the Kraus representation $\varphi(\rho) = \sum_i K^{\dagger}_i \rho K_i$, corresponding to the Markovian master equation~\cite{Lendi}
\begin{align}
    \dot{\rho} &= -i [H,\rho] + \kappa_{\varphi} 
    \sum_i \left( K_i^{\dagger} \rho K_i
    -\frac{1}{2}\left\{ K_i K_i^{\dagger}, \rho \right\} \right),
\end{align}
where $\kappa_{\varphi}$ is the channel's rate. Let the rates $\kappa_{\mathcal{E}}, ~\kappa_{\Phi}$, and $\kappa_{\Psi}$ correspond to the above channels. In the following subsections, we will restrict to two regimes, defined by the magnitude of $\kappa_{\Psi}$ compared to the others: the ``slow'' regime with $\kappa_{\Psi} \ll \kappa_{\mathcal{E}} \ll \kappa_{\Phi}$ and the ``fast'' regime with $\kappa_{\mathcal{E}} \ll \kappa_{\Psi},~ \kappa_{\Phi}$. In Fig.~\ref{fig:quantum_channels}, we visualize these processes in a schematic picture of the structure theory.

\begin{figure}[h!]   
    \centering
    \includegraphics[width=1\linewidth]{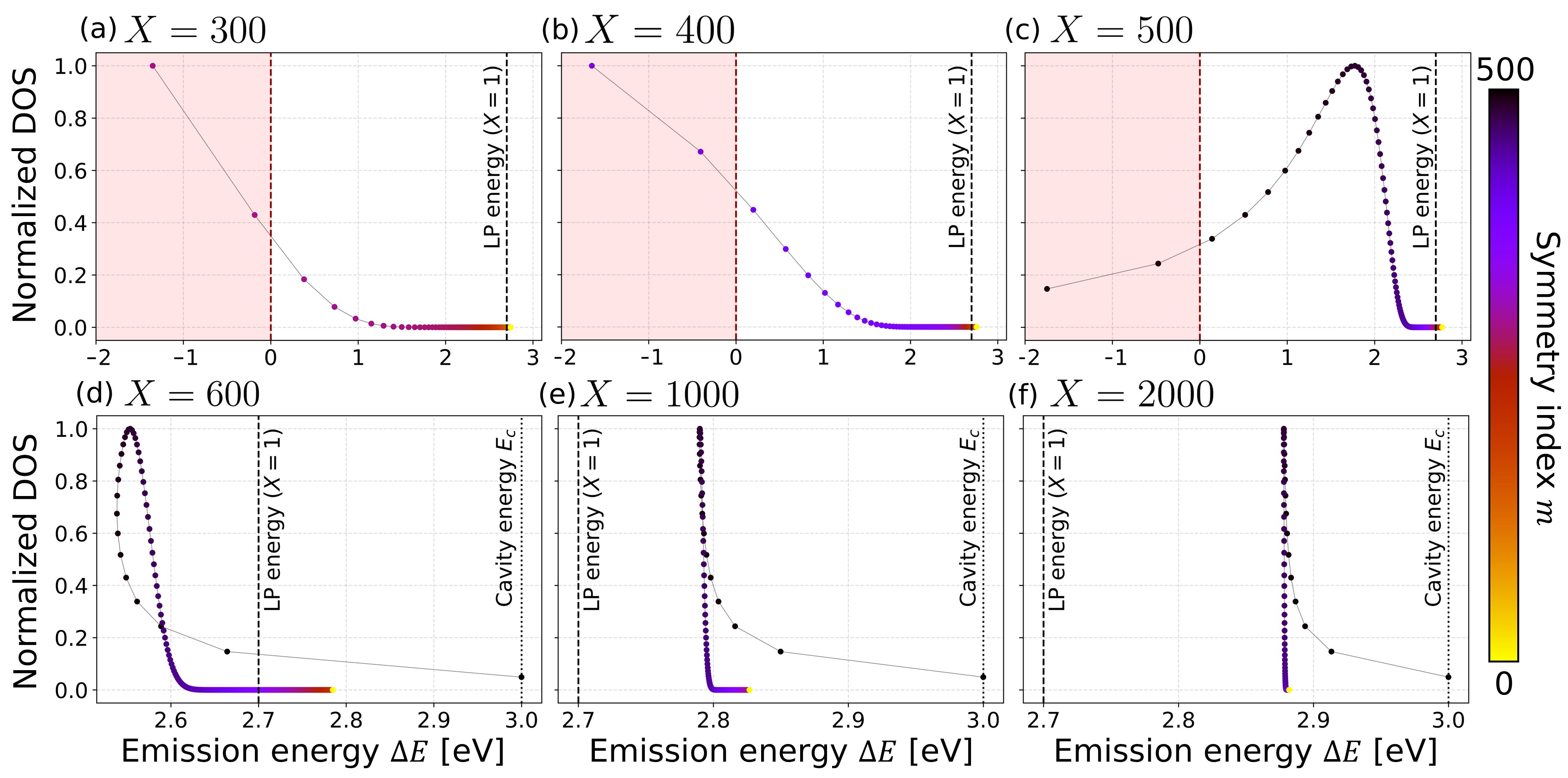}
    \caption{Normalized multiplicities against emission energies for $N=1000$, $g=0.3/\sqrt{N}~\y{eV}$, $E_c = E_{TLS} = 3.0~\y{eV}$ and $X\in\{300, 400, 500, 600, 1000, 2000\}$. Panels (a--c) show how the dark-polariton multiplicities overtake the dark states.
    (d--f) show how for $X>N/2$ the spectrum slowly flips and concentrates around the $m=0$ emission energy. In the negative-energy regions highlighted by red color, the initial state is lower in energy than the final state, and emission cannot occur.}
    \label{fig:emissionN10E3}
\end{figure}

\subsection{Emission energies in the slow dephasing regime}

We are particularly interested in the predictions that our expanded model makes of emission. Assuming fast $m$-conserving relaxation within each irrep, our model predicts emission peaks centered at the energy differences between the irreps' lowest-energy eigenstates.  By Fermi's golden rule~\cite{Chen2026}, the peaks are also proportional to the density of states (DOS), represented here by the discrete multiplicity distribution. Importantly, our model allows for significant numerical optimization when solving these features.

The system can also relax through TLS dephasing~\cite{Siltanen_MH}, which, however, does not necessarily preserve the symmetry index $m$. Let us first consider dephasing-generated rates of internal conversion much slower than emission. In this case, our model allows to estimate both the spectral positions and relative intensities of the (possible) emission peaks. Whether these peaks are actually observable depends on the competing processes that we omit here for simplicity.


Fig.~\ref{fig:emissionN10E3} shows the emission energies $\Delta E = \bra{i}H_{\y{TC}}\ket{i}- \bra{f}H_{\y{TC}}\ket{f} = E^{(X)}_{LMP} - E^{(X-1)}_{LMP}$ between the lowest-energy states of each irrep against normalized multiplicities (DOS) for different values of $X$. Again, the color encodes the symmetry index $m$. Energies adjacent in $m$ are connected by interpolating lines. In all panels, we see that the peaks are extremely dense at low $m$, panel (c) having over half of the peaks above the $X=1$ LP energy, indicated by the dashed line.
Panels (a--c) show that as $X$ grows up to $N/2$, the emission maximum shifts to higher energies, and the total distribution approaches a Poissonian-like form with a maximum red-shifted from the $X=1$ LP energy.

Energy differences below zero appear as well, but they result from the assumption of the initial energy being higher. These sign differences depend on the parameters $N$, $E_c$, and $g$, and negative energies should be interpreted as absorption.

Panels (d--f) show that when $X>N/2$, the high-$m$ emission energies jump near the cavity energy, indicated by the dotted line. When $X$ still keeps increasing, all the peaks shift closer to the cavity, indicating a loss of light--matter hybridization at saturated excitation densities~\cite{Horikiri_2017}. After $X \gtrsim 3N/4$, all of the energies overtake the $X=1$ LP energy, indicating blue shift. At the high-$X$ limit, all the energies collapse on the $m=0$ emission, forming a sharp peak approaching $E_c$. In Appendix~\ref{app:Supplementary_figures}, we present figures for a direct comparison between different excitation manifolds. We also consider energy differences between higher-energy states of the irreps, corresponding to a regime of $m$-conserving rates comparable to emission.

\subsection{Emission energies in the fast dephasing regime}

\begin{figure} 
    \centering
    \includegraphics[width=1\linewidth]{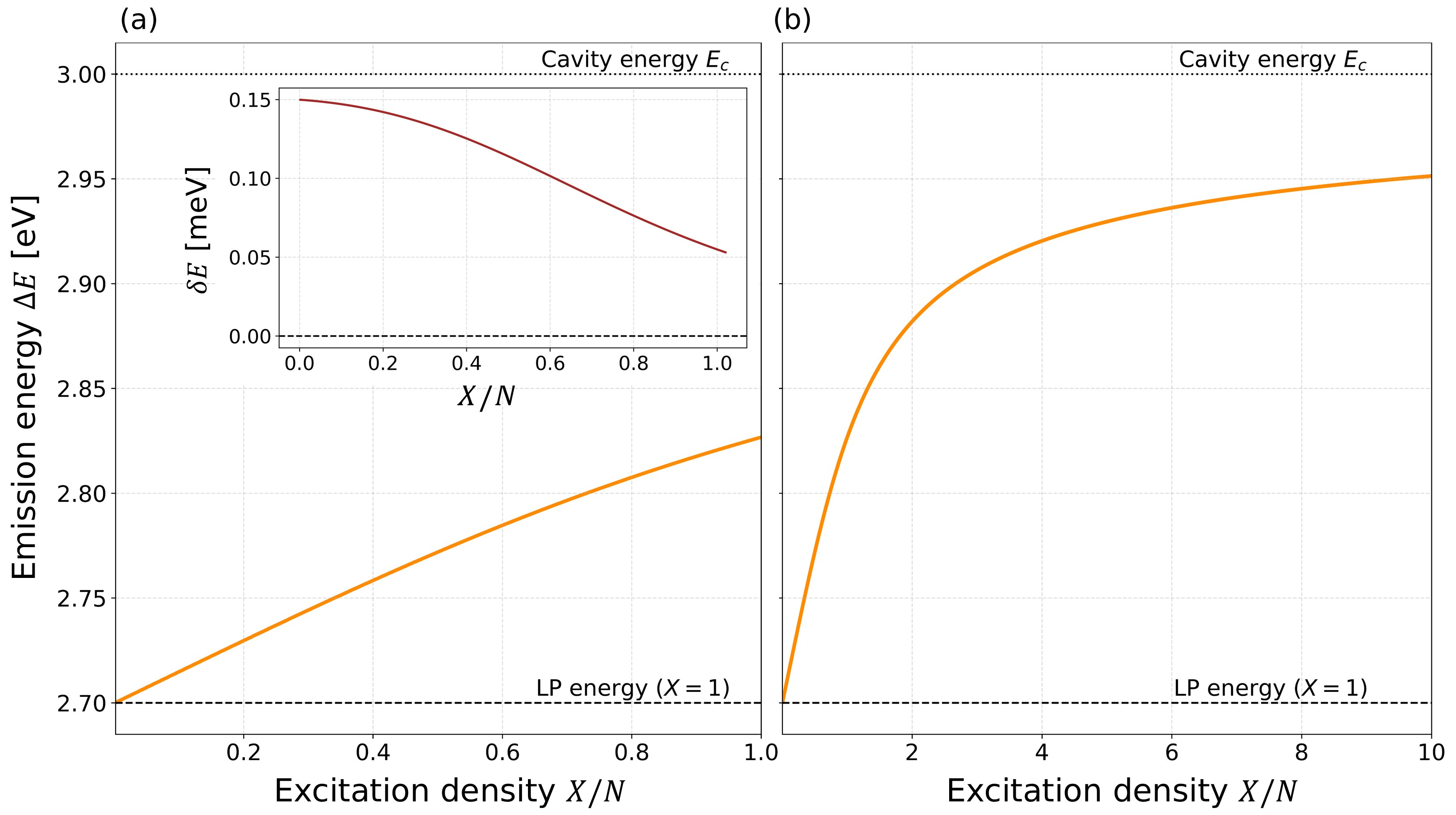}
    \caption{LMP emission energy against total excitation density $X/N$ for $N=1000$. (a) Sublinear blue shift of the LMP emission dominant for fast internal relaxation rates within excitation manifolds. The inset shows the difference between the graph at $N=10^6$ and $N=10^3$, highlighting the small significance of $N$ for the LMP emission. (b) After $X>N$, the LMP emission turns to approach the cavity energy asymptotically in growing $X$.}
    \label{fig:emissionN10E6_g_sqrtN}
\end{figure}

Of greater interest is the regime where internal conversion induced by dephasing dominates, as this is commonly the case~\cite{Siltanen_MH}. Under fast dephasing, the subspaces labeled by $m$ are not closed under the dynamics and the permutational symmetry is perturbed. In this regime, the system first relaxes to the lowest-energy eigenstate of each excitation manifold, and so the LMP states dominate emission.

In Fig.~\ref{fig:emissionN10E6_g_sqrtN}, we have plotted the LMP emission energy as a function of excitation density $X/N$ for $N=1000$. Panel (a) shows a sublinear blue shift of the emission for $0 \leq X \leq N$. This behavior is extremely stable under changing $N$. The inset shows how the emission changes for different $N$, $\delta E := \Delta E (N=10^6) - \Delta E (N=10^3)$, indicating further reduction of this $N$-dependence at higher values of $X$. Panel (b) shows how the emission behavior changes after $X>N$, i.e., when the irrep dimension gets locked  (see Section~\ref{sect:Structure_theory}); the emission peak approaches $E_c$ asymptotically at growing excitation densities. This effect is further strengthened by the collapse of emission energies around the LMP peak (see Fig.~\ref{fig:emissionN10E3}) and the most prominent for high $N$, as the multiplicity approaches a sharp peak at $m=N/2$ for $N \rightarrow \infty$ (see Appendix~\ref{app:High_N_multiplicities}).

Notably, the blue shift in Fig.~\ref{fig:emissionN10E6_g_sqrtN} is similar to previously reported blue shifts in polaritonic systems~\cite{Yagafarov2020,Wei2022}. This blue shift of the classical LP emission peak towards the cavity is expected, though. When the excitation number grows, the TLS part of the system saturates in energy. The irreps reach their maximal dimensions, preventing further TLS excitations by symmetry. Consequently, additional excitations increasingly populate the cavity mode, causing the photonic component of the system to become dominant.
Our model therefore provides a microscopic complement to the mechanism proposed in Refs.~\cite{Yagafarov2020,Wei2022}: the quenching of Rabi splitting due to the saturation of molecular optical transitions, or ``bleaching''.

Finally, the emission results also suggest why the well-established $X=1$ model works so well for modeling observed spectra. For moderate excitation densities, the emission peak stays very close to $E_{LP}$. Small discrepancies from the $X=1$ case might not even be distinguishable due to non-zero linewidths. Higher excitation densities, on the other hand, are naturally suppressed in many applications, e.g., by intermolecular annihilation processes~\cite{Giant-Rabi-Splitting}.


\section*{Discussion}\label{sect:discussion}

In this paper, we derived the structure theory for the TC model with an arbitrary number of TLSs and excitations. We calculated the matrix spectra, eigenstates, and spectral properties of the system, using the representation-theoretic nature of the model to drastically reduce the number of degrees of freedom; the largest matrices computed in this work would have been $\approx 10^{3\cdot10^6}$-dimensional without the provided theory, beyond any other computational method. Still, while the fundamental building blocks of our model remain linear, energy relations receive a highly nonlinear structure at large excitation numbers.

We applied the model to derive selection rules for the dynamical generators of an emitting cavity. We found that at realistic excitation numbers, the radiant regime of energy states (dark polaritons) grows to rival the statistical proportions of the dark states. The selection rules also allowed us to make qualitative predictions of emission. For slow internal dephasing rates, statistical multiplicities of dark polaritons indicate red shift of the emission maximum at lower excitation densities. However, more common faster rates put multipolaritons in a key role, resulting in blue-shifted emission peak at all values of $X$. The predicted blue shift has been experimentally observed~\cite{Yagafarov2020,Wei2022}, although establishing a direct connection between our theory and experiment requires further investigation and is left for future work. Furthermore, the multipolariton emission peak was found to be near the $X=1$ LP emission, showing why the widely used $X=1$ model fits to experimental data despite its simplicity. Finally, the structure was shown to saturate at excitation numbers $\mathcal{O}(N)$, leading to an asymptotic cavity-like behavior.

The analysis was carried out under zero detuning and uniform coupling, considering only a single cavity mode. This allowed us to obtain analytical results that can also be extended to less restrictive assumptions and other parameter regimes using perturbative approaches. While it is clear that more extensions of the model are needed, it is this simplicity that allows for a clear first understanding of these regimes of higher energy.

In general, our work unveils the rich structure of the uncomputably large energy space of the full TC model while also taming its size. The model creates an understandable general picture of the full quantum mechanical system, covering a vast range of applications due to its fundamental nature. It sets the stage for experimental comparison with the predicted emissive behavior. In addition to the spectral study of this work, the presented model also works as a tool for studying processes involving many excitations, such as annihilation processes. This exciting regime is fundamentally invisible to the few-excitation models, pinpointing a way forward for further fundamental understanding.

\paragraph{Data availability}
The codes for generating the data and the figures are available at \url{https://github.com/LMD-UTU/SC_w_X_excitations}.

\paragraph{Acknowledgments}
This project has received funding from the Research Council of Finland project ``X-SHIELD'' (decision number 369819) and the European Research Council (ERC) under the European Union's Horizon 2020 research and innovation program (grant agreement number 948260). Views and opinions expressed are, however, those of the authors only and do not necessarily reflect those of the European Union. Neither the European Union nor the granting authority can be held responsible for them. OS acknowledges financial support from the Research Council of Finland under PROFI 7 -- Strengthening the Profiling of Universities, on Sustainable Materials and Manufacturing (``SUSMAT'', decision number 352727).

\paragraph{Author contributions}
AP carried out the analysis and wrote the first draft. OS conceptualized the work and supervised it together with KL and KSD. KSD was responsible for funding acquisition, project administration, and resources. All authors discussed the contents and participated in reviewing and editing.

\bibliographystyle{unsrtnat}
\bibliography{References}

@article{Campos-Gonzalez-Angulo_2,
    author = {Campos-Gonzales-Angulo, J. A. and Yuen-Zhou, J.},
    year = {2022},
    journal = {Journal of Chemical Physics},
    pages = {},
    title = {Generalization of the Tavis-Cummings model for multi-level anharmonic systems: insights on the second excitation manifold},
    doi = {10.48550/arXiv.2202.01433}
}

@article{Dicke,
  title = {Coherence in Spontaneous Radiation Processes},
  author = {Dicke, R. H.},
  journal = {Physical Review},
  volume = {93},
  issue = {1},
  pages = {99--110},
  numpages = {0},
  year = {1954},
  publisher = {American Physical Society},
  doi = {10.1103/PhysRev.93.99},
}

@article{Tavis-Cummings,
  title = {Exact Solution for an $N$-Molecule---Radiation-Field Hamiltonian},
  author = {Tavis, M. and Cummings, F. W.},
  journal = {Physical Review},
  volume = {170},
  issue = {2},
  pages = {379--384},
  numpages = {0},
  year = {1968},
  publisher = {American Physical Society},
  doi = {10.1103/PhysRev.170.379},
}

@article{Lednev2024,
archivePrefix = {arXiv},
arxivId = {2305.13171},
author = {Lednev, M. and Garc{\'{i}}a-Vidal, F. J. and Feist, J.},
doi = {10.1103/PhysRevLett.132.106902},
eprint = {2305.13171},
issn = {10797114},
journal = {Physical Review Letters},
number = {10},
pages = {106902},
pmid = {38518335},
publisher = {American Physical Society},
title = {{Lindblad Master Equation Capable of Describing Hybrid Quantum Systems in the Ultrastrong Coupling Regime}},
volume = {132},
year = {2024}
}

@article{Sandik2024,
author = {Sandik, G. and Feist, J. and Garc{\'{i}}a-Vidal, F. J. and Schwartz, T.},
doi = {10.1038/s41563-024-01962-5},
issn = {1476-4660},
journal = {Nature Materials 2024 24:3},
number = {3},
pages = {344--355},
pmid = {39122930},
publisher = {Nature Publishing Group},
title = {{Cavity-enhanced energy transport in molecular systems}},
volume = {24},
year = {2024}
}

@article{Dutta2024,
author = {Dutta, A. and Tiainen, V. and Sokolovskii, I. and Duarte, L. and Marke{\v{s}}evi{\'{c}}, N. and Morozov, D. and Qureshi, H. A. and Pikker, S. and Groenhof, G. and Toppari, J. J.},
doi = {10.1038/s41467-024-50532-5},
issn = {2041-1723},
journal = {Nature Communications 2024 15:1},
number = {1},
pages = {6600--},
pmid = {39097575},
publisher = {Nature Publishing Group},
title = {{Thermal disorder prevents the suppression of ultra-fast photochemistry in the strong light-matter coupling regime}},
volume = {15},
year = {2024}
}

@article{Sanvitto2016,
author = {Sanvitto, D. and K{\'{e}}na-Cohen, S.},
doi = {10.1038/nmat4668},
issn = {1476-1122},
journal = {Nature Materials},
number = {10},
pages = {1061--1073},
publisher = {Nature Publishing Group},
title = {{The road towards polaritonic devices}},
volume = {15},
year = {2016}
}

@article{Bhuyan2023,
author = {Bhuyan, R. and Mony, J. and Kotov, O. and Castellanos, G. W. and {G{\'{o}}mez Rivas}, J. and Shegai, T. O. and B{\"{o}}rjesson, K.},
doi = {10.1021/acs.chemrev.2c00895},
issn = {0009-2665},
journal = {Chemical Reviews},
number = {18},
pages = {10877--10919},
pmid = {37683254},
publisher = {American Chemical Society},
title = {{The Rise and Current Status of Polaritonic Photochemistry and Photophysics}},
volume = {123},
year = {2023}
}

@article{Abdelmagid2025,
archivePrefix = {arXiv},
arxivId = {2412.06741},
author = {Abdelmagid, A. G. and Qiao, Z. and Coenegracht, B. and Yu, G. and Qureshi, H. A. and Anthopoulos, T. D. and Gasparini, N. and Daskalakis, K. S.},
doi = {10.1002/adom.202501727},
eprint = {2412.06741},
issn = {2195-1071},
journal = {Advanced Optical Materials},
number = {28},
pages = {1--7},
title = {{Polaritons in Non‐Fullerene Acceptors for High Responsivity Angle‐Independent Organic Narrowband Infrared Photodiodes}},
volume = {13},
year = {2025}
}

@book{Klimov_Chumakov,
    author = {Klimov, A. B. and Chumakov, S. M.},
    publisher = {John Wiley \& Sons, Ltd},
    isbn = {9783527624003},
    title = {A Group‐Theoretical Approach to Quantum Optics},
    booktitle = {A Group‐Theoretical Approach to Quantum Optics},
    chapter = {},
    pages = {I-IX},
    doi = {https://doi.org/10.1002/9783527624003.fmatter},
    eprint = {https://onlinelibrary.wiley.com/doi/pdf/10.1002/9783527624003.fmatter},
    year = {2009}
    }

@article{Campos-Gonzalez-Angulo_1,
    doi = {10.1088/1367-2630/ac00d7},
    year = {2021},
    publisher = {IOP Publishing},
    volume = {23},
    number = {6},
    pages = {063081},
    author = {Campos-Gonzalez-Angulo, J. A. and Ribeiro, R. F. and Yuen-Zhou, J.},
    title = {Generalization of the Tavis–Cummings model for multi-level anharmonic systems},
    journal = {New Journal of Physics},
}

@book{Fulton1991RepresentationTA,
  title={Representation Theory: A First Course},
  booktitle = {Representation Theory: A First Course},
  publisher = {New York, Springer},
  author={Fulton, W. and Harris, J. W.},
  year={1991},
  doi = {10.1007/978-1-4612-0979-9}
}

@book{Sagan,
    title ={The symmetric Group: Representations, Combinatorial Algorithms, and Symmetric Functions (Graduate Texts in Mathematics)},
    booktitle = {The symmetric Group: Representations, Combinatorial Algorithms, and Symmetric Functions (Graduate Texts in Mathematics)},
    publisher ={New York, Springer},
    author = {Sagan, B.},
    year = {2013},
    doi = {10.1007/978-1-4757-6804-6}
}

@book{Fulton_1996_Young_tableaux, 
    place={Cambridge}, 
    series={London Mathematical Society Student Texts}, 
    title={Young Tableaux: With Applications to Representation Theory and Geometry}, 
    publisher={Cambridge University Press}, 
    author={Fulton, W.},
    year={1996}, 
    collection={London Mathematical Society Student Texts},
    doi = {10.1017/CBO9780511626241}
}

@article{Skrypnyk_2018,
    doi = {10.1088/1751-8121/aa94af},
    year = {2017},
    publisher = {IOP Publishing},
    volume = {51},
    number = {1},
    pages = {015204},
    author = {Skrypnyk, T.},
    title = {Modified n-level, n - 1-mode Tavis–Cummings model and algebraic Bethe ansatz},
    journal = {Journal of Physics A: Mathematical and Theoretical},
}

@article{PAN200594,
    title = {Exact solutions of an extended Dicke model},
    journal = {Physics Letters A},
    volume = {341},
    number = {1},
    pages = {94-100},
    year = {2005},
    issn = {0375-9601},
    doi = {https://doi.org/10.1016/j.physleta.2005.05.005},
    author = {Pan, F. and Wang, T. and Pan, J. and Li Y. and Draayer, J. P.},
}

@article{Sanchez_CUT-E,
    author = {Pérez-Sánchez, J. B. and Koner, A. and Raghavan-Chitra, S. and Yuen-Zhou, J.},
    title = {CUT-E as a 1/N expansion for multiscale molecular polariton dynamics},
    journal = {The Journal of Chemical Physics},
    volume = {162},
    number = {6},
    pages = {064101},
    year = {2025},
    issn = {0021-9606},
    doi = {10.1063/5.0244452},
    eprint = {https://pubs.aip.org/aip/jcp/article-pdf/doi/10.1063/5.0244452/20386730/064101_1_5.0244452.pdf},
}

@article{Siltanen_AOM,
author = {Siltanen, O. and Luoma, K. and Musser, A. J. and Daskalakis, K. S.},
title = {Enhancing the Efficiency of Polariton OLEDs in and Beyond the Single-Excitation Subspace},
journal = {Advanced Optical Materials},
volume = {13},
number = {12},
pages = {2403046},
doi = {https://doi.org/10.1002/adom.202403046},
eprint = {https://advanced.onlinelibrary.wiley.com/doi/pdf/10.1002/adom.202403046},
year = {2025}
}

@article{Borges_Dark_Polariton,
    author = {Borges, L. and Schnappinger, T. and Kowalewski, M.},
    title = {Impact of Dark Polariton States on Collective Strong Light–Matter Coupling in Molecules},
    journal = {The Journal of Physical Chemistry Letters},
    volume = {16},
    number = {31},
    pages = {7807-7815},
    year = {2025},
    doi = {10.1021/acs.jpclett.5c01480},
    eprint = {https://doi.org/10.1021/acs.jpclett.5c01480}
}

@article{Abdelmagid_Gruyter,
    title = {Identifying the origin of delayed electroluminescence in a polariton organic light-emitting diode},
    author = {Abdelmagid, A. G. and Qureshi, H. A. and Papachatzakis, M. A. and Siltanen, O. and Kumar, M. and Ashokan, A. and Salman, S. and Luoma, K. and Daskalakis, K. S.},
    pages = {2565--2573},
    volume = {13},
    number = {14},
    journal = {Nanophotonics},
    doi = {doi:10.1515/nanoph-2023-0587},
    year = {2024},
    lastchecked = {2026-07-21}
}

@article{Giant-Rabi-Splitting,
    author = {Qureshi, H. A. and Papachatzakis, M. A. and Abdelmagid, A. G. and Salomäki, M. and Mäkilä, E. and Tuomi, O. and Siltanen, O. and Daskalakis, K. S.},
    title = {Giant Rabi Splitting and Polariton Photoluminescence in an all Solution-Deposited Dielectric Microcavity},
    journal = {Advanced Optical Materials},
    volume = {13},
    number = {16},
    pages = {2500155},
    doi = {https://doi.org/10.1002/adom.202500155},
    eprint = {https://advanced.onlinelibrary.wiley.com/doi/pdf/10.1002/adom.202500155},
    year = {2025}
}

@article{Embrace_the_darkness,
    author = {Khazanov, T. and Gunasekaran, S. and George, A. and Lomlu, R. and Mukherjee, S. and Musser, A. J.},
    title = {Embrace the darkness: An experimental perspective on organic exciton–polaritons},
    journal = {Chemical Physics Reviews},
    volume = {4},
    number = {4},
    pages = {041305},
    year = {2023},
    issn = {2688-4070},
    doi = {10.1063/5.0168948},
    eprint = {https://pubs.aip.org/aip/cpr/article-pdf/doi/10.1063/5.0168948/18207413/041305_1_5.0168948.pdf},
}

@Article{Qureshi2026,
author={Qureshi, H. A.
and Lyyra, H.
and Korkeam{\"a}ki, A.
and Tuomi, O.
and Moilanen, A. J.
and Daskalakis, K. S.},
title={A fully solution-processed organic microcavity laser in the strong light-matter coupling regime},
journal={Nature Communications},
year={2026},
volume={17},
number={1},
pages={8280},
issn={2041-1723},
doi={10.1038/s41467-026-75118-1},
}

@article{Polaritonic_quantum_matter,
    author = {Basov, D. N. and Asenjo-Garcia, A. and Schuck, J. P. and Zhu, X. and Rubio, A. and Cavalleri, A. and Delor, M. and Fogler, M. M. and Liu, M.},
    title = {Polaritonic quantum matter},
    journal = {Nanophotonics},
    volume = {14},
    number = {23},
    pages = {3723-3760},
    doi = {https://doi.org/10.1515/nanoph-2025-0001},
    eprint = {https://onlinelibrary.wiley.com/doi/pdf/10.1515/nanoph-2025-0001},
    year = {2025}
}

@article{Siltanen_MH,
    author = {Siltanen, O. and Luoma, K. and Daskalakis, K. S.},
    title = {Impact of light–matter coupling strength on the efficiency of microcavity OLEDs: a unified quantum master equation approach},
    journal = {Materials Horizons},
    volume = {13},
    number = {7},
    pages = {3343-3354},
    year = {2026},
    issn = {2051-6347},
    doi = {10.1039/d5mh01958c},
    eprint = {https://pubs.rsc.org/mh/article-pdf/13/7/3343/10452152/d5mh01958c.pdf},
}

@article{Molecular_polaritons_review,
    author = {Xiang, B. and Xiong, W.},
    title = {Molecular Polaritons for Chemistry, Photonics and Quantum Technologies},
    journal = {Chemical Reviews},
    volume = {124},
    number = {5},
    pages = {2512-2552},
    year = {2024},
    doi = {10.1021/acs.chemrev.3c00662},
    eprint = { https://doi.org/10.1021/acs.chemrev.3c00662}
}

@Article{Byrnes2014,
    author={Byrnes, T.
    and Kim, N. Y.
    and Yamamoto, Y.},
    title={Exciton--polariton condensates},
    journal={Nature Physics},
    year={2014},
    volume={10},
    number={11},
    pages={803-813},
    doi={10.1038/nphys3143},
}

@Article{Hymas2026,
    author={Hymas, K.
    and Muir, J. B.
    and Tibben, D.
    and van Embden, J.
    and Hirai, T.
    and Dunn, C. J.
    and G{\'o}mez, D. E.
    and Hutchison, J. A.
    and Smith, T. A.
    and Quach, J. Q.},
    title={Superextensive electrical power from a quantum battery},
    journal={Light: Science {\&} Applications},
    year={2026},
    volume={15},
    number={1},
    pages={168},
    issn={2047-7538},
    doi={10.1038/s41377-026-02240-6},
}

@Article{Mischok2024,
    author={Mischok, A.},
    title={Polaritons light up future displays},
    journal={Light: Science {\&} Applications},
    year={2024},
    volume={13},
    number={1},
    pages={302},
    issn={2047-7538},
    doi={10.1038/s41377-024-01647-3},
}

@Article{Pandya2021,
    author={Pandya, R. and et al.},
    title={Microcavity-like exciton-polaritons can be the primary photoexcitation in bare organic semiconductors},
    journal={Nature Communications},
    year={2021},
    volume={12},
    number={1},
    pages={6519},
    issn={2041-1723},
    doi={10.1038/s41467-021-26617-w},
}

@article{Tibben_2025,
  title = {Extending the Self-Discharge Time of Dicke Quantum Batteries Using Molecular Triplets},
  author = {Tibben, D. J. and Della Gaspera, E. and van Embden, J. and Reineck, P. and Quach, J. Q. and Campaioli, F. and G\'omez, D. E.},
  journal = {PRX Energy},
  volume = {4},
  issue = {2},
  pages = {023012},
  numpages = {10},
  year = {2025},
  publisher = {American Physical Society},
  doi = {10.1103/bhyh-53np},
}

@article{Haroche_experiment,
  title = {Observation of Self-Induced Rabi Oscillations in Two-Level Atoms Excited Inside a Resonant Cavity: The Ringing Regime of Superradiance},
  author = {Kaluzny, Y. and Goy, P. and Gross, M. and Raimond, J. M. and Haroche, S.},
  journal = {Physical Review Letters},
  volume = {51},
  issue = {13},
  pages = {1175--1178},
  numpages = {0},
  year = {1983},
  publisher = {American Physical Society},
  doi = {10.1103/PhysRevLett.51.1175},
}

@article{Braak,
    author = {Braak, D.},
    title = {Symmetries in the Quantum Rabi Model},
    journal = {Symmetry},
    year = {2019},
    volume = {11},
    issue = {1259},
    doi = {10.3390/sym11101259}
}

@article{Ribeiro_nonlinear,
  title = {Enhanced optical nonlinearities under collective strong light-matter coupling},
  author = {Ribeiro, R. F. and Campos-Gonzalez-Angulo, J. A. and Giebink, N. C. and Xiong, W. and Yuen-Zhou, J.},
  journal = {Physical Review A},
  volume = {103},
  issue = {6},
  pages = {063111},
  numpages = {24},
  year = {2021},
  publisher = {American Physical Society},
  doi = {10.1103/PhysRevA.103.063111},
}

@Article{Yagafarov2020,
author={Yagafarov, T.
and Sannikov, D.
and Zasedatelev, A.
and Georgiou, K.
and Baranikov, A.
and Kyriienko, O.
and Shelykh, I.
and Gai, L.
and Shen, Z.
and Lidzey, D.
and Lagoudakis, P.},
title={Mechanisms of blueshifts in organic polariton condensates},
journal={Communications Physics},
year={2020},
volume={3},
number={1},
pages={18},
doi={10.1038/s42005-019-0278-6},
}

@article{Vandaele_2017,
doi = {10.1088/1751-8121/aa5bc2},
year = {2017},
publisher = {IOP Publishing},
volume = {50},
number = {11},
pages = {114002},
author = {Vandaele, E. R. J. and Arvanitidis, A. and Ceulemans, A.},
title = {The quantization of the Rabi Hamiltonian},
journal = {Journal of Physics A: Mathematical and Theoretical}
}

@article{Mandal,
    author = {Mandal, A. and Taylor, M. A. D. and Weight, B. M. and Koessler, E. R. and Li, X. and Huo, P.},
    title = {Theoretical Advances in Polariton Chemistry and Molecular Cavity Quantum Electrodynamics},
    journal = {Chemical Reviews},
    volume = {123},
    number = {16},
    pages = {9786-9879},
    year = {2023},
    doi = {10.1021/acs.chemrev.2c00855},
    eprint = {https://doi.org/10.1021/acs.chemrev.2c00855}
}

@ARTICLE{Fregoni2022-qg,
  title     = "Theoretical challenges in polaritonic chemistry",
  author    = "Fregoni, J. and Garcia-Vidal, F. J. and Feist,
               J.",
  journal   = "ACS Photonics",
  publisher = "American Chemical Society (ACS)",
  volume    =  9,
  number    =  4,
  pages     = "1096--1107",
  year      =  2022,
  language  = "en",
  doi       = "10.1021/acsphotonics.1c01749"
}

@article{Weight_ab_initio_methods,
author = {Weight, B. M. and Huo, P.},
title = {Ab Initio Approaches to Simulate Molecular Polaritons and Quantum Dynamics},
journal = {WIREs Computational Molecular Science},
volume = {15},
number = {4},
pages = {e70039},
doi = {https://doi.org/10.1111/wcms.70039},
eprint = {https://wires.onlinelibrary.wiley.com/doi/pdf/10.1111/wcms.70039},
note = {e70039 CMS-1031.R1},
year = {2025}
}

@ARTICLE{DelPo2020-yx,
  title     = "Polariton transitions in femtosecond transient absorption
               studies of ultrastrong light-molecule coupling",
  author    = "DelPo, C. A. and Kudisch, B. and Park, K. H. and
               Khan, S. and Fassioli, F. and Fausti, D.
               and Rand, B. P. and Scholes, G. D.",
  journal   = "Journal of Physical Chemistry Letters",
  publisher = "American Chemical Society (ACS)",
  volume    =  11,
  number    =  7,
  pages     = "2667--2674",
  year      =  2020,
  copyright = "https://creativecommons.org/licenses/by-nc-nd/4.0/",
  language  = "en",
  doi       = "10.1021/acs.jpclett.0c00247"
}

@article{Sanchez-Barquilla,
    author = {Sánchez-Barquilla, M. and Fernández-Domínguez, A. I. and Feist, J. and García-Vidal, F. J.},
    title = {A Theoretical Perspective on Molecular Polaritonics},
    journal = {ACS Photonics},
    volume = {9},
    number = {6},
    pages = {1830-1841},
    year = {2022},
    doi = {10.1021/acsphotonics.2c00048},
    eprint = { https://doi.org/10.1021/acsphotonics.2c00048}
}

@ARTICLE{Mandal2023-jz,
  title     = "Theoretical advances in polariton chemistry and molecular cavity
               quantum electrodynamics",
  author    = "Mandal, A. and Taylor, M. A. D. and Weight, B. M. and
               Koessler, E. R. and Li, X. and Huo, P.",
  journal   = "Chemical Reviews",
  publisher = "American Chemical Society (ACS)",
  volume    =  123,
  number    =  16,
  pages     = "9786--9879",
  year      =  2023,
  copyright = "https://creativecommons.org/licenses/by/4.0/",
  language  = "en",
  doi       = "10.1021/acs.chemrev.2c00855"
}

@book{Stevens,
    author = {Stevens, J.},
    title = {Schur-Weyl duality},
    publisher = {Department of Mathematics, University of Chicago},
    year = {2016}
}

@article{Cauchy_interlacing,
    author = {Said, K.},
    journal = {arXiv:1603.04151},
    year = {2016},
    pages = {},
    title = {The Cauchy interlace theorem for symmetrizable matrices},
    doi = {10.48550/arXiv.1603.04151}
}

@book{Horn_Johnson_1985, 
    place={Cambridge}, 
    title={Matrix Analysis}, 
    publisher={Cambridge University Press}, 
    author={Horn, R. A. and Johnson, C. R.}, 
    year={1985}
}

@article{Ruseckas_SSA,
    author = {Ruseckas, A. and Ribierre, J. C. and Shaw, P. E. and Staton, S. V. and Burn, P. L. and Samuel, I. D. W.},
    title = {Singlet energy transfer and singlet-singlet annihilation in light-emitting blends of organic semiconductors},
    journal = {Applied Physics Letters},
    volume = {95},
    number = {18},
    pages = {183305},
    year = {2009},
    issn = {0003-6951},
    doi = {10.1063/1.3253422},
    eprint = {https://pubs.aip.org/aip/apl/article-pdf/doi/10.1063/1.3253422/14423687/183305_1_online.pdf},
}

@book{Breuer_Petruccione,
    author = {Breuer, H. P. and Petruccione, F.},
    title = {The Theory of Open Quantum Systems},
    publisher = {Oxford University Press},
    year = {2007},
    isbn = {9780199213900},
    doi = {10.1093/acprof:oso/9780199213900.001.0001},
}

@article{Scala_Piilo,
  title = {Microscopic derivation of the Jaynes-Cummings model with cavity losses},
  author = {Scala, M. and Militello, B. and Messina, A. and Piilo, J. and Maniscalco, S.},
  journal = {Physical Review A},
  volume = {75},
  issue = {1},
  pages = {013811},
  numpages = {8},
  year = {2007},
  publisher = {American Physical Society},
  doi = {10.1103/PhysRevA.75.013811},
}

@Article{Chen2026,
    author={Chen, J.
    and Huang, S.
    and Ji, Y.
    and Schumacher, G. L.
    and Tsidilkovski, A.
    and Schuckert, A.
    and T. Assump{\c{c}}{\~a}o, G. G.
    and Navon, N.},
    title={Emergence of Fermi's golden rule in a quantum many-body system},
    journal={Nature Physics},
    year={2026},
    issn={1745-2481},
    doi={10.1038/s41567-026-03316-1},
}

@article{Stembridge,
    author = {Stembridge, J. R.},
    title = {{On the eigenvalues of representations of reflection groups and wreath products.}},
    volume = {140},
    journal = {Pacific Journal of Mathematics},
    number = {2},
    publisher = {Pacific Journal of Mathematics, A Non-profit Corporation},
    pages = {353 -- 396},
    year = {1989},
    doi = {10.2140/pjm.1989.140.353}
}

@Article{Wei2022,
    author={Wei, M.
    and Verstraelen, W.
    and Orfanakis, K.
    and Ruseckas, A.
    and Liew, T. C. H.
    and Samuel, I. D. W.
    and Turnbull, G. A.
    and Ohadi, H.},
    title={Optically trapped room temperature polariton condensate in an organic semiconductor},
    journal={Nature Communications},
    year={2022},
    volume={13},
    number={1},
    pages={7191},
    issn={2041-1723},
    doi={10.1038/s41467-022-34440-0}
}

@book{Abramowitz,
    author = {Abramowitz, M. and Stegun I. A.},
    title = {Handbook of Mathematical Functions},
    publisher = {Dover},
    year = {1965},
    doi = {10.2307/2314682}
}

@book{Srednicki_2007, 
    place={Cambridge}, 
    title={Quantum Field Theory}, 
    publisher={Cambridge University Press}, 
    author={Srednicki, M.}, 
    year={2007},
    doi = {10.1017/CBO9780511813917}
}

@article{Horikiri_2017,
  title = {Highly excited exciton-polariton condensates},
  author = {Horikiri, T. and Byrnes, T. and Kusudo, K. and Ishida, N. and Matsuo, Y. and Shikano, Y. and L\"offler, A. and H\"ofling, S. and Forchel, A. and Yamamoto, Y.},
  journal = {Physical Review B},
  volume = {95},
  issue = {24},
  pages = {245122},
  numpages = {5},
  year = {2017},
  publisher = {American Physical Society},
  doi = {10.1103/PhysRevB.95.245122},
}

@book{Lendi,
    author = {Alicki, R. and Lendi, K.},
    publisher = {Springer Berlin, Heidelberg},
    isbn = {978-3-540-70860-5},
    title = {Quantum Dynamical Semigroups and Applications},
    booktitle = {Quantum Dynamical Semigroups and Applications},
    chapter = {},
    pages = {I-IX},
    doi = {https://doi.org/10.1007/3-540-70861-8},
    year = {2007},
    }

\onecolumn\newpage
\appendix

\noindent\Large{\textbf{APPENDIX}}
\normalsize

\section{Young tableaux and hook lengths}\label{app:Young_tableaux}

A \emph{Young diagram} is a matrix of left-justified rows of $N$ cells, with given shape $\lambda$ defined by a partition of $N$, $\lambda = (n_1, n_2, \dots,n_k)$, where $\sum_{i=1}^k n_i = N$. It is common to list the $k$ \textit{parts} $n_i$ of the partition in decreasing order. For example, the Young diagram corresponding to $N=9$ and $\lambda = (4,2,2,1)$ is in \emph{English notation}
\begin{center}
\ytableausetup{mathmode,centertableaux}
\begin{ytableau}
~ &~  &~  &~  \\
~ & ~  \\
~& ~ \\
~  
\end{ytableau}.
\end{center}
Two different definitions are used, \emph{English} or \emph{French} notation, differing only in the order of the parts $n_i$, i.e, the same diagram in French notation would be
\begin{center}
\ytableausetup{mathmode,centertableaux}
\begin{ytableau}
~   \\
~ & ~  \\
~& ~ \\
~ &~  &~  &~  
\end{ytableau}.
\end{center}
We fix to use English notation in this work. We can further label the cells of a Young diagram with an ordered set of $N$ symbols to get a \emph{Young tableau}. This labeling is in a sense arbitrary, and can be fixed to be the integers $[1,N]$. The \emph{canonical labeling} is the one with natural ordering of the numbers from the top left row-wise, for example for the diagram above
\begin{center}
\ytableausetup{mathmode,centertableaux}
\begin{ytableau}
1 &2  &3  &4  \\
5 & 6  \\
7& 8 \\
9
\end{ytableau}.
\end{center}

A tableau is called a \emph{standard Young tableau} (SYT) if the labeling on each row is strictly increasing, and similarly a \emph{semistandard Young tableau} (SSYT) if the labeling is non-decreasing. The notion of the SSYT allows us to consider numbers reappearing in the diagram, which can happen naturally in some applications~\cite{Fulton_1996_Young_tableaux}. An SSYT can be equivalently described through strictly increasing labels moving down along columns. The canonically labeled tableau is an SYT, and an example of an SSYT would be
\begin{center}
\ytableausetup{mathmode,centertableaux}
\begin{ytableau}
1 &2  &2  &4  \\
3 & 4  \\
5 & 5 \\
7
\end{ytableau}.
\end{center}

Following the convention of (weakly) decreasing parts $n_i$ of partition $\lambda$, the number of cells on each row must be non-increasing. For example for $N=4$, all the diagrams considered are
\begin{center}
\ytableausetup{mathmode,centertableaux}
\begin{ytableau}
~ & ~& ~&~  \\
\end{ytableau},~~
\begin{ytableau}
~&~&~ \\
~
\end{ytableau},~~
\begin{ytableau}
~&~ \\
~&~
\end{ytableau},~~
\begin{ytableau}
~&~ \\
~ \\
~
\end{ytableau},~~
\begin{ytableau}
~ \\
~ \\
~ \\
~
\end{ytableau}.
\end{center}

The number of SYTs with $N$ cells is known to correspond to the sequence of integers known as the involution numbers~\cite{Stembridge}. The number of SSYTs related to a \emph{specific shape} $\lambda$ is also known, and it is given by the famous hook length formula~\cite{Sagan}. The \textit{hook length} of a cell $h(i,j)$ is the number of cells directly below and to the right of the cell at $i$th row and $j$th column, counting the cell itself. This shape forms a ``hook'' to the right, giving the name. This quantity is then used for the \emph{hook length formula}
\begin{align}
    \dim{S^{\lambda}} &= \frac{N!}{\prod h(i,j)},
\end{align}
where $N$ is the number of cells and $S^{\lambda}$ is an irrep of $S_N$~\cite{Fulton_1996_Young_tableaux}. As mentioned above, this quantity is equal to the number of SSYT relating to shape $\lambda$ of $N$, giving an alternative combinatorial method to calculate irrep dimensions (Section~\ref{sect:Structure_theory}).

Consider, e.g., arbitrary $N$ and $\lambda = (N-2,2)$, corresponding to the diagram
\begin{center}
\ytableausetup{mathmode,centertableaux}
\begin{ytableau}
~ &~  & \none[\cdots]  &~  \\
~ & ~ 
\end{ytableau},
\end{center}
with $N-2$ cells on the first row. In  this two-row case, the hook lengths can be seen to be
\begin{center}
\ytableausetup{mathmode, boxsize=2em}
\begin{ytableau}
\scriptstyle N-1 & \scriptstyle N-2 & \scriptstyle N-4 & \none[\cdots] & 2 & 1 \\
2 & 1 
\end{ytableau},
\end{center}
where the values of the omitted cells decrease by one repeatedly. We see that the product $\prod h(i,j) = 2 \cdot (N-1)!/(N-3)$, such that the hook length formula gives $\dim{S^{\lambda}} = \frac{1}{2}N(N-3)$, which is the multiplicity of the $m=2$ subspace for $N$ TLSs of Section~\ref{sect:Structure_theory}. 

\section{Eigenstate properties}\label{app:Eigenstate_properties}

\subsection{Symmetric spectrum}\label{app:Symmetric_spectrum}

Consider the $n \times n$ dimensional reduced interaction matrix $A$, defined through $H_{TC} = H_0 + gA$. Let $\mathbf{v} \in \mathbb{C}^n$ be an eigenvector, $A \mathbf{v} = \lambda \mathbf{v}$, where $\lambda \in \mathbb{R}$ since $A$ is real and symmetric. Then $A$ and the matrix $D = D^{-1} = \y{diag}(\dots,-1,1,-1,1)_{n \times n}$ satisfy 
\begin{align}
    DAD^{-1} = -A &\Leftrightarrow DA = -AD, \\
    \Rightarrow DH_{\y{TC}}D^{-1} &= H_0 - gA.
\end{align}
Therefore $D\mathbf{v}$ must also be an eigenvector, with eigenvalue $-\lambda$, as $A(D\mathbf{v}) = -DA\mathbf{v} = -\lambda (D\mathbf{v})$. The spectrum of $A$ is therefore symmetric about zero, and as for zero detuning $H_0 = XE_c \identity_{n \times n} = XE_{TLS} \identity_{n \times n}$, the true spectrum is symmetric about $XE_c = XE_{TLS}$.

\subsection{Average excitation contents}\label{app:Average_excitation_contents}

The TC model eigenstates form a basis for each $\mathcal{H}_m^{(X)}$, and so the \textit{average of the average excitation contents over $\mathcal{H}_m^{(X)}$} is a basis-invariant quantity. We can therefore calculate the average photonic/TLS excitation content of the $\mathcal{H}_m^{(X)}$ eigenstates using the bare symmetric basis $\Vec{\alpha}_k = (\delta_k^l \alpha_l)_{X-m+1}$. Noting that the component $\alpha_k$ refers to photonic content $k$, we find
\begin{align}
    \langle n_c \rangle_{\mathcal{H}_m^{(X)}} 
    &= \frac{1}{X-m+1}\sum_{k=0}^{X-m} 
    \Vec{\alpha}_k^{\dagger}(\hat{n}_c) \Vec{\alpha}_k \\
    &= \frac{1}{X-m+1} \sum_{k=0}^{X-m} k \\
    &= \frac{X-m}{2},
\end{align}
using the sum of the first $X-m$ integers. By definition $\langle n_{TLS} \rangle_{\mathcal{H}_m^{(X)}} = X- \langle n_c \rangle_{\mathcal{H}_m^{(X)}}$, and so
\begin{align}
    \langle n_{TLS} \rangle_{\mathcal{H}_m^{(X)}} 
    &= \frac{X+m}{2}.
\end{align}

\subsection{Parity alternation}\label{app:Parity_alternation}

We can show that the asymptotic sign parities of the $m=0$ irrep must alternate in increasing eigenenergies. We will prove the result in two parts, separating it into even and odd dimension of the Hilbert space.

\subsubsection{Even-dimensional case}

Let the asymptotic reduced interaction matrix $\Tilde{A~} \in \mathcal{M}_{2n}(\mathbb{R})$, and the set $\{ \lambda_i\}_{i=1}^{2n}$ be its eigenvalues in increasing order. This is sensible, as $\Tilde{A~}$ has a real spectrum due to it being Hermitian. By the results of Section~\textbf{\ref{sect:eigenstates_selection_rules_and_emission}}, we know that the matrix decomposes into block form by the eigenvalues $\pm 1$ of the exchange matrix $J$, as
\begin{align}\label{eqn:parity_sub-blocks}
    \Tilde{A~} &= \Tilde{A~}^+ \oplus \Tilde{A~}^-,
\end{align}
where the two sub-blocks $\Tilde{A~}^+, \Tilde{A~}^- \in \mathcal{M}_{n}(\mathbb{R})$. This must be, as $\y{tr}[J_{2n}]=0$. Consider then the natural eigenbasis of $J$
\begin{align}\label{eqn:exchange_basis_even}
    \mathcal{B} &= \mathcal{B}^+ \cup \mathcal{B}^- \notag \\
    &=\{ \mathbf{w}_1^+,\dots,\mathbf{w}_n^+ \} \cup
    \{ \mathbf{w}_1^-,\dots,\mathbf{w}_n^- \},
\end{align}
where $\mathbf{w}_i^\pm = \frac{1}{\sqrt{2}}(\mathbf{e}_i \pm \mathbf{e}_{2n+1-i})$, and $\mathbf{e}_i \in \mathbb{R}^{2n}$ is the unit vector along the $i$th component in the symmetric basis. The matrix elements of $\Tilde{A~}$ in the symmetric basis are $(\Tilde{A~})_{ij} = c_k \delta_1^{|i-j|}\delta_k^{\min{\{i,j\}}}$, with $c_k = \sqrt{k(X+1-k)}$ for $X+1=2n$ and $k =1,2,\dots, 2n-1$. Therefore, the matrix elements in the basis Eq.~\eqref{eqn:exchange_basis_even} are
\begin{align}
    \Tilde{A~}\mathbf{w}_i^\pm 
    &= \frac{1}{\sqrt{2}}(\Tilde{A~}\mathbf{e}_i
    +\Tilde{A~}\mathbf{e}_{2n+1-i}) \notag \\
    &=\frac{1}{\sqrt{2}}(c_{i-1}\mathbf{e}_{i-1}
    + c_i\mathbf{e}_{i+1}
    \pm c_{2n-i}\mathbf{e}_{2n-i}
    \pm c_{2n+1-i}\mathbf{e}_{2n+2-i}) \notag \\
    &=\frac{1}{\sqrt{2}}(c_{i-1}\mathbf{e}_{i-1}
    + c_i\mathbf{e}_{i+1}
    \pm c_{i}\mathbf{e}_{2n-i}
    \pm c_{i-1}\mathbf{e}_{2n+2-i}) \notag \\
    &= c_{i-1}\mathbf{w}_{i-1}^\pm + c_i\mathbf{w}_{i+1}^\pm,
\end{align}
where we used the persymmetric property $c_{2n-k}=c_k$. The edge cases result in
\begin{align}
    \Tilde{A~}\mathbf{w}_1^\pm 
    &=c_1 \mathbf{w}_2^{\pm}, \\
    \Tilde{A~}\mathbf{w}_n^\pm
    &= c_{n-1}\mathbf{w}_{n-1}^\pm \pm c_n \mathbf{w}_n^\pm,
\end{align}
showing that the two parity sub-blocks are equivalent, up to a parity-dependent diagonal term $\Tilde{A~}^+ = \Tilde{A~}^- + c_n \mathbf{e}_{nn}$ in the basis Eq.~\eqref{eqn:exchange_basis_even}, where $\mathbf{e}_{ij}$ is the matrix unit.

Since $\Tilde{A~}$ is Hermitian and we can write $c_n \mathbf{e}_{nn} = (\sqrt{c_n}\mathbf{e}_n)\cdot (\sqrt{c_n}\mathbf{e}_n)^{\dagger}$, we can invoke Corollary~4.3.9 of Ref.~\cite{Horn_Johnson_1985} to deduce
\begin{align}\label{eqn:parity_alternation}
    \lambda_i(\Tilde{A~}^-) 
    \leq \lambda_i(\Tilde{A~}^+)
    \leq \lambda_i(\Tilde{A~}^-),
\end{align}
where $\lambda_i(M)$ is the $i$th eigenvalue of the matrix $M$ in increasing order. But since $\Tilde{A~}= \Tilde{A~}^+ \oplus \Tilde{A~}^-$ has simple eigenvalues, the equalities cannot hold, and the parities must alternate.

\subsubsection{Odd-dimensional case}

\noindent
Let $\Tilde{A~} \in \mathcal{M}_{2n+1}(\mathbb{R})$, and the set $\{\lambda_i\}_{i=1}^{2n+1}$ be its eigenvalues in increasing order. Since $\y{tr}[J_{2n+1}]=1$, we find the block decomposition by parities Eq.~\eqref{eqn:parity_sub-blocks} with $\Tilde{A~}^+ \in \mathcal{M}_{2n+1}(\mathbb{R})$ and $\Tilde{A~}^- \in \mathcal{M}_{2n}(\mathbb{R})$. Consider the basis
\begin{align}\label{eqn:exchange_basis_odd}
    \mathcal{B} &= \mathcal{B}^+ \cup \mathcal{B}^- \notag \\
    &=\{ \mathbf{w}_1^+,\dots,\mathbf{w}_n^+, \mathbf{w}_{n+1} \} \cup
    \{ \mathbf{w}_1^-,\dots,\mathbf{w}_n^- \},
\end{align}
where $\mathbf{w}_i^\pm = \frac{1}{\sqrt{2}}(\mathbf{e_i} \pm \mathbf{e}_{2n+2-i})$, and $\mathbf{w}_{n+1} := 2^{1/4}\mathbf{e}_{n+1}$ with $\mathbf{e}_i \in \mathbb{R}^{2n+1}$.

The matrix elements of $\Tilde{A~}$ in the symmetric basis are $(\Tilde{A~})_{ij} = c_k \delta_1^{|i-j|}\delta_k^{\min{\{i,j\}}}$, with $c_k = \sqrt{k(X+1-k)}$ for $X+1=2n+1$ and $k =1,2,\dots, 2n$. Similarly to the even-dimensional case, we find
\begin{align}
    \Tilde{A~} \mathbf{w}_i^{\pm}
    &= \frac{1}{\sqrt{2}} ( c_{i-1}\mathbf{e}_{i-1}
    +c_i\mathbf{e}_{i+1}
    \pm c_i \mathbf{e}_{2n+1-i}
    \pm c_{i-1} \mathbf{e}_{2n+3-i}) \notag \\
    &= c_{i-1}\mathbf{w}_{i-1}^\pm + c_i \mathbf{w}_{i+1}^\pm,
\end{align}
with edge cases
\begin{align}
    \Tilde{A~}\mathbf{w}_1^\pm 
    &=c_1 \mathbf{w}_2^{\pm}, \\
    \Tilde{A~}\mathbf{w}_n^\pm
    &= c_{n-1}\mathbf{w}_{n-1}^\pm +
    \begin{cases}
        2^{1/4}c_n\mathbf{w}_{n+1}, & (+) \\
        0, & (-)
    \end{cases}\\
    \Tilde{A~} \mathbf{w}_{n+1} 
    &= 2^{1/4}c_n\mathbf{w}_n^+,
\end{align}
where we used $c_{2n+1-k} = c_k$. We see that both matrices $\Tilde{A~}^\pm$ stay tridiagonal, and that $\Tilde{A~}^-$ is the first principal submatrix of $\Tilde{A~}^+$. Since $\Tilde{A~}^+$ is also real and symmetric, the eigenvalues must alternate like Eq.~\eqref{eqn:parity_alternation} with strict inequalities according to the Cauchy interlacing theorem~\cite{Cauchy_interlacing} and simplicity of the eigenvalues.

\subsection{Extremal eigenstates}\label{app:Extremal_eigenstates}

A calculated guess suggests that the maximal eigenstates (states with the highest eigenvalue) of the matrix $\Tilde{A~} := \lim_{N\rightarrow \infty} \frac{1}{\sqrt{N}}A$ are given by the components
\begin{align*}
    \xi_k := \ket{E^{(X,0)}_X}_k = \sqrt{\frac{1}{2^X}\binom{X}{X+1-k}}, \tag{Re.~(\ref{eqn:Extremal_Eigenstate_Components})}
\end{align*}
for $m=0$, $1\leq k\leq X+1$ as shown in the main work. This conjecture is seen to be true as follows.

According to Table~\ref{BareElements2}, the matrix elements of $\Tilde{A~}$ are $(\Tilde{A~})_{ij} = \Tilde{c}_n \delta_1^{|i-j|}\delta_n^{\min{\{i,j\}}}$, where $\Tilde{c}_n = \sqrt{n(X+1-n)}$ for $1\leq n\leq X$. The action of $\Tilde{A~}$ is then found component-wise to be: for the first component
\begin{align}
    \xi_1 \mapsto c_1 \xi_2  &= \sqrt{X}\sqrt{\frac{1}{2^X}\binom{X}{X-1}} \notag \\
    &= \frac{X}{2^X} = X\xi_1,
\end{align}
and for the last component
\begin{align}
    \xi_{X+1} \mapsto c_X \xi_X &= \sqrt{X(X+1-X)}\sqrt{\frac{1}{2^X}\binom{X}{X+1-X}} \notag \\
    &= \frac{X}{2^X} = X\xi_{X+1}.
\end{align}
For the intermediate components, one gets
\begin{align}
    \xi_k &\mapsto c_{k-1}\xi_{k-1}+c_k\xi_{k+1} \notag \\
    &= \frac{1}{\sqrt{2^X}}\left(
    \sqrt{(k-1)(X+2-k)\binom{X}{X+2-k}}+\sqrt{k(X+1-k)\binom{X}{X-k}}
    \right) \notag \\
    &= \frac{1}{\sqrt{2^X}} \left(
    (k-1)\sqrt{\binom{X}{X+1-k}} + (X+1-k)\sqrt{\binom{X}{X+1-k}}
    \right) \notag \\
    &= \sqrt{\frac{1}{2^X}\binom{X}{X+1-k}}X = X\xi_k,
\end{align}
showing that the components $\xi_k$ give the eigenvector with eigenvalue $X$. On the other hand, the Geršgorin disc theorem gives an upper bound for the absolute values of the eigenvalues, as the highest sum of the matrix elements on each row~\cite{Horn_Johnson_1985}. This sum is $ B := \Tilde{c}_k + \Tilde{c}_{k-1}$ for the $k$th row, where $\Tilde{c}_k \equiv 0$ for $k<1$ and $k>X$. Inputting the values of $\Tilde{c}$ this can be further evaluated to $B\leq \frac{X}{2}+\frac{X}{2} = X$, giving an upper bound of $X$. Therefore the state given by components $\xi_k$ is the maximal eigenstate.

By the calculations of Appendix~\ref{app:Symmetric_spectrum}, the eigenstate with the eigenvalue $-X$ is given by $D\Vec{\xi}$ with components
\begin{align}
    (D)_{kk} \xi_k &= \xi_k (-1)^{k-X-1},
\end{align}
for $1\leq k\leq X+1$. As the ordering of the eigenstates is not affected by $\lim_{N\rightarrow \infty}$, this defines the minimal-energy eigenstate of each excitation manifold.

\subsection{Transition probabilities for \texorpdfstring{$m=0$} {TEXT}}
\label{app:Transition_probabilities}

We can use the results of Appendix~\ref{app:Extremal_eigenstates} to prove Eq.~\eqref{eqn:Transition_Probability} for the limit $N \rightarrow \infty$,
\begin{align*}
    |\bra{E^{(X-1)}_{LMP}}\hat{a}\ket{E^{(X)}_{LMP}}|^2
    &= \bra{E^{(X)}_{LMP}}\hat{n}\ket{E^{(X)}_{LMP}}
    =\frac{X}{2},
\end{align*}
i.e., that for the LMP states the photonic content and transition probabilities squared are equal, evaluated $X/2$. Since the binomial coefficient is symmetric $\binom{X}{k} = \binom{X}{X-k}$, we may write
\begin{align}
    \bra{E^{(X)}_{LMP}}  \hat{n} \ket{E^{(X)}_{LMP}} 
    &= \left ( \sum_{k=0}^X \xi_k^*\sqrt{\binom{N}{k}^{-1}}
    \sum_{\{n_i \}_{i=0}^k} \bra{e_{\{n_i \}}}\bra{X-k} \right ) \notag \\
    &~~~ \times \left (\sum_{k'=0}^X \xi_{k'}\sqrt{\binom{N}{k'}^{-1}}
    \sum_{\{n_j \}_{j=0}^{k'}} \ket{e_{\{n_j \}}}\ket{X-k'} \right ) \notag \\
    &= \sum_{k=0}^X |\xi_k|^2(X-k) \notag \\
    &= \sum_{k=0}^X \frac{1}{2^X}\binom{X}{k}(X-k) = \frac{X}{2},
\end{align}
where $\ket{e_{\{n_i \}}}= \ket{e_{n_0} e_{n_1} \cdots e_{n_{k-1}} e_{n_k}}$ in increasing order of $n_i$. 

The transition probability can be found similarly,
 \begin{align}
     |\bra{E^{(X-1)}_{LMP}}\hat{a}\ket{E^{(X)}_{LMP}}|^2 
     &= \left| \left( \sum_{k'=0}^{X-1} \xi_k^{(X-1)}\sqrt{\binom{N}{k'}^{-1}}
     \sum_{ \{n_j \}_{j=0}^{k'}} \bra{e_{\{n_j \} }}\bra{X-k'}\right) \right. \notag \\
     &~~~ \times \hat{a}
     \left.  \left( \sum_{k=0}^{X-1} \xi_k^{(X-1)}\sqrt{\binom{N}{k}^{-1}}
     \sum_{ \{n_i \}_{i=0}^{k}} \ket{e_{\{n_i \} }}\ket{X-k}\right) \right|^2 \notag \\
     &= \left| \left( \sum_{k,k'=0}^X \xi_k^{(X-1)} \xi_k^{(X)}\sqrt{X-k'}
     \sqrt{ \binom{N}{k'}^{-1} \binom{N}{k}^{-1}}\right) \right. \notag \\
     &~~~ \times \left. \left( \sum_{\{n_j \},\{n_i \}}
     \braket{e_{\{n_j \}}}{e_{\{n_i \}}} \braket{X-k'}{X-k} \right) \right|^2 \notag \\
     &= \left| \sum_{k=0}^X \xi_k^{(X-1)} \xi_k^{(X)}\sqrt{X-k} \right|^2 \notag \\
     &=\frac{1}{2^{2X-1}} \left| \sum_{k=0}^X \sqrt{\binom{X}{k}\binom{X-1}{k}(X-k)}
     \right|^2 \notag \\
     &= \frac{1}{2^{2X-1}}\frac{1}{X} \left| \sum_{k=0}^X \binom{X}{k}(X-k) \right|^2\notag \\
     &= \frac{1}{2^{2X-1}}\frac{1}{X} \left| 2^{X-1} X \right|^2 \notag \\
     &= \frac{X}{2},
 \end{align}
proving our claim.

\section{High-\texorpdfstring{$N$}{TEXT} multiplicities}\label{app:High_N_multiplicities}

Let us inspect what happens to the multiplicity distribution Eq.~\eqref{eqn:irrep_dimension} at the high-$N$ limit.
Because of the hypergeometric nature of $d(N,m)$, it is not obvious how the peak will move, or even if more peaks will appear. We then need to inspect this uniqueness as well. We search for the peak of the multiplicity distribution with respect to $m$ depending on $N$. Let us first analytically continue the formula with gamma functions
\begin{align}
    d(N,m) = \frac{\Gamma(N+1)(N-2m+1)}{\Gamma(m+1)\Gamma(N-m+2)},
\end{align}
by the well known relation for integers $n! = \Gamma(n+1)$. The function is then differentiable in $\mathbb{C}$. We want to find the extremal point(s)
\begin{align}
    \frac{\partial}{\partial m}d(N,m) &= 0,~~0\leq m\leq N/2\text{.}
\end{align}
To do this, we need to be able to calculate the derivative of the gamma function. This problem can be solved using the \emph{digamma function}, defined as
\begin{align}
    \psi(z) &:=\frac{d}{dz}\mathrm{ln}~\Gamma(z) = \frac{\Gamma^{'}(z)}{\Gamma(z)},
\end{align}
where $z \in \mathbb{C}$ and $\Gamma^{'}(z) = \frac{d}{dz} \Gamma(z)$. We may then write
\begin{align}
    \frac{d}{dz}\Gamma(z) &= \Gamma(z)\psi(z),
\end{align}
defining the derivative of the gamma function~\cite{Abramowitz}. The calculation then proceeds as follows:
\begin{align}\label{eqn:high_N_multiplicity_step1}
    0 &= \frac{\partial}{\partial m} d(N,m) \notag \\
    &= \Gamma(N+1) \frac{\partial}{\partial m}\left(
    \frac{N-2m+1}{\Gamma(m+1)\Gamma(N-m+2)}  \right) \notag \\
    &= \Gamma(N+1) \left[ \frac{-2}{\Gamma(m+1)\Gamma(N-m+2} \right. \notag \\
    &~~~+ \left. (N-2m+1)\frac{\partial}{\partial m}
    \left( \frac{1}{\Gamma(m+1)\Gamma(N-m+2)} \right)   \right] \notag \\
    \Leftrightarrow 2 &= (N-2m+1) \frac{f'(m)}{f(m)},
\end{align}
where we defined $f(m) := 1/\Gamma(m+1)\Gamma(N-m+2)$ and $f'(m) = \frac{\partial}{\partial m}f(m)$. Now since $\frac{\partial}{\partial x} \y{ln}(x) = 1/x$, we may write
\begin{align}
    \frac{1}{f(m)}\frac{\partial}{\partial m}f(m) 
    &= \frac{\partial}{\partial m}\left[ \y{ln} \left( \frac{1}{\Gamma(m+1)\Gamma(N-m+2)}
    \right) \right] \notag \\
    &= \frac{\partial}{\partial m} \big[ \y{ln}(1)-\y{ln}(\Gamma(m+1))- \y{ln}(\Gamma(N-2m+1))
    \big] \notag \\
    &= \frac{\partial}{\partial m} \big[-\y{ln}(\Gamma(m+1))- \y{ln}(\Gamma(N-2m+1))
    \big],
\end{align}
by standard differentiation rules. Now, by definition $\psi(m) = \frac{\partial}{\partial m} \y{ln}~\Gamma(m)$ and $\frac{\partial}{\partial m} [-\y{ln}~\Gamma(N+(-m)+2)] = \frac{\partial}{\partial (-m)}(\y{ln}~\Gamma(N-m+2)) = \psi(N-m+2)$, so we get from Eq.~\eqref{eqn:high_N_multiplicity_step1}
\begin{align}\label{eqn:high_N_multiplicity_step2}
    \frac{2}{N-2m+1} &= \psi(N-m+2)-\psi(m+1),
\end{align}
the solution(s) of which (in terms of $m$) give the extremal point(s) of $d(N,m)$. This can be seen to have only one solution in the interval $0\leq m\leq N/2$ as follows. Take the second derivative of Eq.~\eqref{eqn:high_N_multiplicity_step2},
\begin{align}
    -\psi'(N-m+2)-\psi'(m+1)-\frac{4}{(N-2m+1)^2},
\end{align}
where all terms (before the minus signs) are positive, such that the second derivative is strictly negative,
\begin{align}
        -\psi'(N-m+2)-\psi'(m+1)-\frac{4}{(N-2m+1)^2} &< 0.
\end{align}
This means that the first derivative is strictly decreasing. Furthermore, since $\psi(z)$ grows rapidly as $N$ grows, $d(N,m)$ will be sharply centered near its solution. Then the edge cases of our original condition for large $N$ can be approximated as shown next. For $m=0$, we can write
\begin{align}
    \psi(N+2)- \psi(1) -\frac{2}{N+1} &\approx \psi(N+2) > 0,
\end{align}
so the first derivative starts as positive. For $m=N/2$, we get
\begin{align}
    & \psi\left(\frac{N}{2}+2\right) -\psi\left(\frac{N}{2}+1\right)-2 \notag \\ 
    &= \frac{d}{dN}\left(
    \y{ln}~\Gamma\left(\frac{N}{2}+2\right)-\y{ln}~\Gamma\left(\frac{N}{2}+1\right) \right)-2 \notag \\
    &= \frac{d}{dN}\y{ln}~\left( \frac{\Gamma(N/2+2)}{\Gamma(N/2+1)}\right) -2 \notag \\
    &\xrightarrow{N \rightarrow \infty} -2 <0,
\end{align}
meaning that the sign of the first derivative changes. But since the first derivative is strictly decreasing, the equation Eq.~\eqref{eqn:high_N_multiplicity_step2} must have only one solution.

Now the solution of the extremal point condition cannot be solved analytically from Eq.~\eqref{eqn:high_N_multiplicity_step2}. However, we can try to find the asymptotic solution at $N \rightarrow \infty$. We do this by considering the leading term of the Laurent series~\cite{Srednicki_2007}
\begin{align}
    \psi(z) &= \y{ln}(z) - \mathcal{O}(z^{-1}),
\end{align}
cutting off terms $\mathcal{O}(z^{-1})$. At high $N$ we then get approximately
\begin{align}
    \frac{2}{N-2m} &= \y{ln}(N-m)-\y{ln}(m) \notag \\
    &= \y{ln}\left(\frac{N-m}{m}\right),
\end{align}
where the constant terms were ignored. For fixed $m$ at the high-$N$ limit, the left-hand side goes to zero and we get
\begin{align}
    0 &= \y{ln}\left(\frac{N-m}{m}\right) \notag \\
    \Rightarrow  1 &= \frac{N-m}{m} \notag \\
    \Leftrightarrow m &= \frac{N}{2}
\end{align}
in the leading term regime. This means that at high numbers of TLSs, the multiplicity becomes sharply peaked at the irrep $V_m^{(X)}$ with $m = N/2$. The magnitude of the numbers appearing from $d(N,m)$ becomes extremely large very quickly, limiting numerical work due to memory limitations. However, normalized numerical calculations can still be done and are considered in the main work.

\section{Supplementary figures}\label{app:Supplementary_figures}

\begin{figure}[h]
    \centering
    \includegraphics[width=1\linewidth]{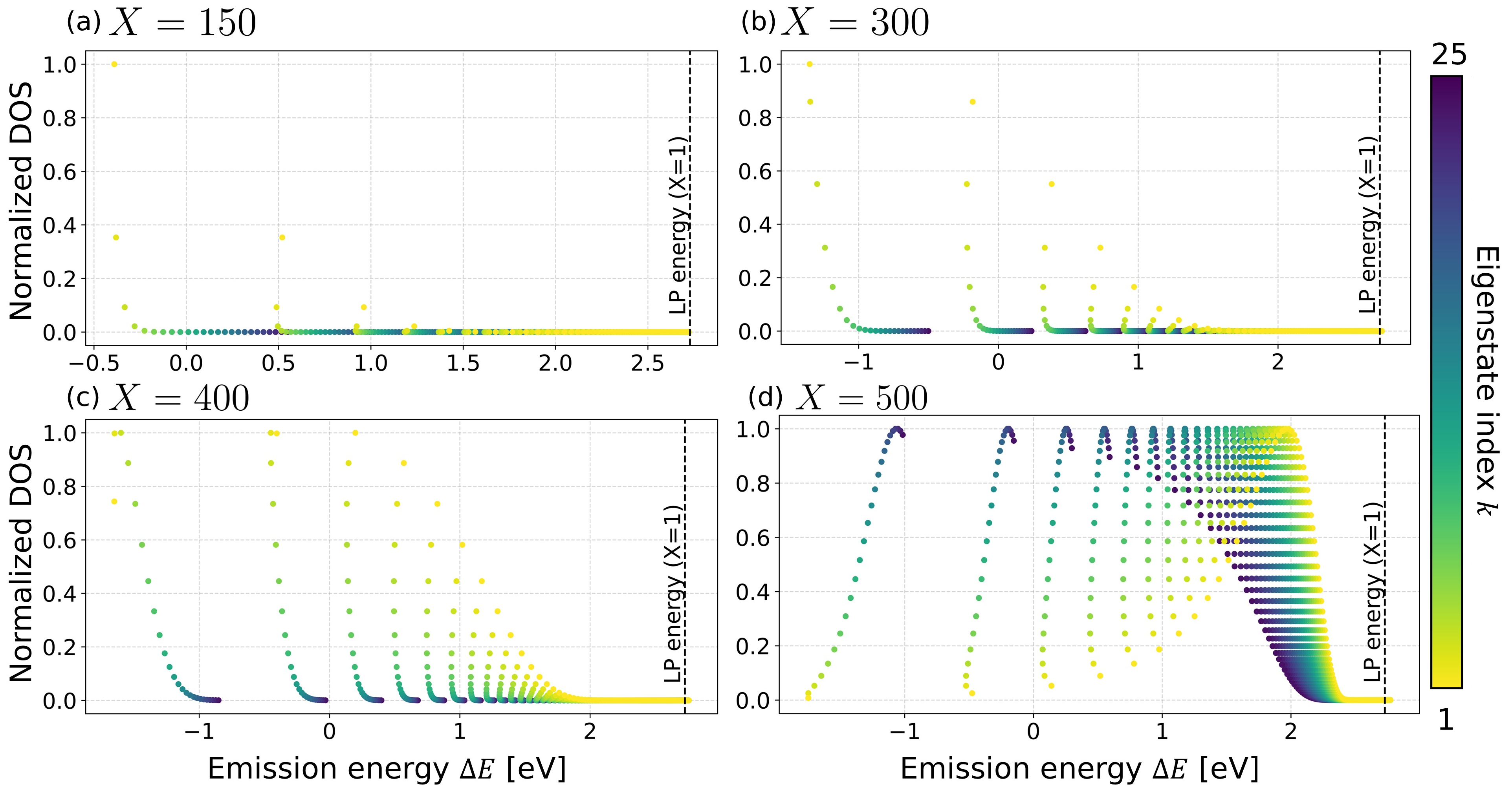}
    \caption{Normalized multiplicities against emission energies for $N=1000$, $g=0.3/\sqrt{N}~\y{eV}$, $E_c = E_{TLS} = 3.0~\y{eV}$, $X\in\{150, 300, 400, 500\}$, and eigenstate index $1 \leq k \leq 25$. Panels (a--c) show how energies for $k>1$ are dominated by the $k=1$ emission. Panel (d) shows how the $k>1$ energies become comparable to $k=1$, and how their distribution follows the results of the main work, with a delay in $X$.}
    \label{fig:multi_k_emission}
\end{figure}

For this appendix, we fix $N=1000, g = 0.3/\sqrt{N}~\y{eV}$, and $E_c = T_{TLS} = 3.0~\y{eV}$.
Let the integer $k$ label eigenstates corresponding to a given pair $(X,m)$ in increasing order of energy. In Fig.~\ref{fig:multi_k_emission}, we present the slow-regime emission distributions for $X \in \{150, 300, 400, 500\}$, and with the eigenstate index $k\in[1,25]$ color-coded. The points for $k=1$ follow the results of Section~\ref{sect:eigenstates_selection_rules_and_emission}. We see that the states with $k>1$ follow similar trends as $k=1$, only with a delay in $X$. 

It is important to note that for $k>1$, the plotted energy-DOS distribution becomes less justified as a true emission distribution. This is because in reality, different values of $k$ are not equally probable to emit. The system naturally tends to minimize energy, making lower values of $k$ more probable.

\begin{figure}[h]
    \centering
    \includegraphics[width=1\linewidth]{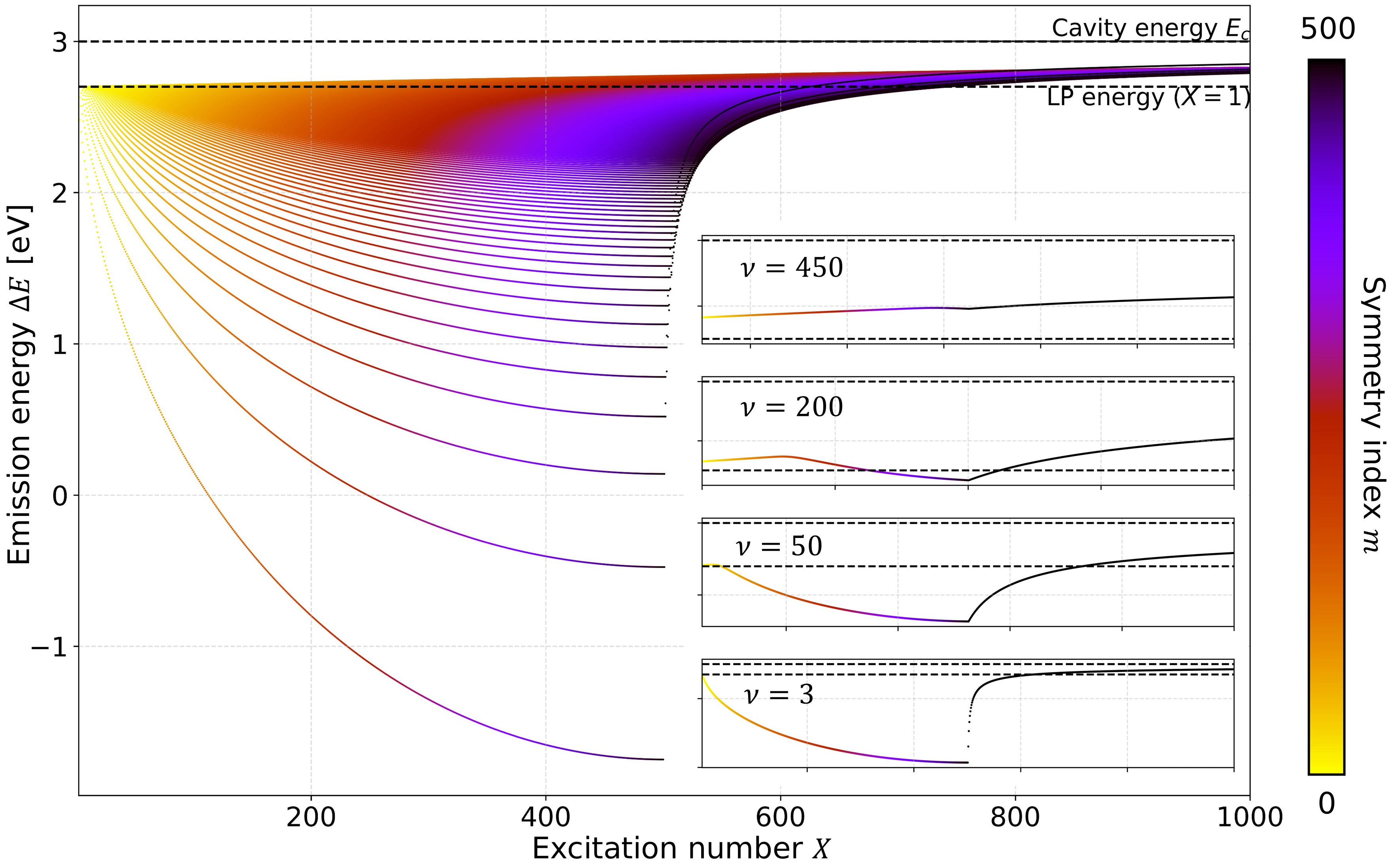}
    \caption{Emission energy for $N=1000$, $g=0.3/\sqrt{N}~\y{eV}$, $E_c = E_{TLS} = 3.0~\y{eV}$, and $k=1$ against excitation number $X$. The distribution spreads out until $X>N/2$, after which it collapses to converge towards $E_c$. The different curves can be labeled by $\nu:= X-m$. The insets show these curves separately for $\nu\in\{3, 50, 200, 450\}$, where we see the collapsing nature of the distribution weaken for higher  $\nu$.}
    \label{fig:multi_manifold_emission}
\end{figure}

In Fig.~\ref{fig:multi_manifold_emission}, we plot the emission energy from the lowest-energy states in each irrep as a function of excitation number $X$. The color encodes the symmetry index $m$. The emission distribution spreads out for growing $X\leq N/2$ but collapses to converge towards the cavity energy $E_c$ after $X>N/2$. This is explained by the locking of the irrep dimensions after $X>N-m$ seen in Section~\ref{sect:Structure_theory}. As seen in Section~\ref{sect:eigenstates_selection_rules_and_emission}, the full distribution becomes blue-shifted compared to the LP energy after $X \gtrsim 3N/4$, with the effect only increasing for higher $X$.

We see distinguishable curves forming in Fig.~\ref{fig:multi_manifold_emission}. These correspond to the energy ordering, indexed by the integer $ \nu := X-m = \dim{\mathcal{H}_m^{(X)}}-1$ for $X<N-m$. When $X\geq N-m$, the relevant integer is the value of $m$ for which the dimensional locking first occurs. This forms a trajectory on tables such as Table~\ref{dimensiontableN9}, following a diagonal with constant integer coefficient until reaching a value $m$ for which the dimension has locked. The trajectory then goes down with fixed $m$. The insets of Fig.~\ref{fig:multi_manifold_emission} show these curves separately for $\nu \in \{ 3, 50, 200, 450\}$. We see that the ``kink'' near $X=N/2$ straightens out for large $\nu$. The highest energy for each $X$ corresponds to the inspection of $m=0$ in Section~\ref{sect:eigenstates_selection_rules_and_emission}. Also note that the low-$m$ emission peaks do not disappear at high $X$, but the distribution becomes so dense that they become indistinguishable from other peaks.

\end{document}